\documentclass[aps,prb,twocolumn,showpacs,preprintnumbers,amsmath,amssymb,superscriptaddress]{revtex4}%

\usepackage{graphicx}
\usepackage{dcolumn}
\usepackage{bm}
\usepackage{color}

\begin{document}

\preprint{preprint(\today)}

\title{Muon-spin rotation and magnetization studies of chemical and hydrostatic pressure effects in EuFe$_{2}$(As$_{1-x}$P$_{x}$)$_{2}$}

\author{Z.~Guguchia}
\email{zurabgug@physik.uzh.ch} \affiliation{Physik-Institut der
Universit\"{a}t Z\"{u}rich, Winterthurerstrasse 190, CH-8057
Z\"{u}rich, Switzerland}


\author{A.~Shengelaya}
\affiliation{Department of Physics, Tbilisi State University,
Chavchavadze 3, GE-0128 Tbilisi, Georgia}

\author{A.~Maisuradze}
\affiliation{Physik-Institut der Universit\"{a}t Z\"{u}rich,
Winterthurerstrasse 190, CH-8057 Z\"{u}rich, Switzerland}
\affiliation{Laboratory for Muon Spin Spectroscopy, Paul Scherrer Institute, CH-5232
Villigen PSI, Switzerland}

\author{L.~Howald}
\affiliation{Physik-Institut der Universit\"{a}t Z\"{u}rich,
Winterthurerstrasse 190, CH-8057 Z\"{u}rich, Switzerland}

\author{Z.~Bukowski}
\altaffiliation{Current address: Institute of Low Temperature and Structure
Research, Polish Academy of Sciences, 50-422 Wroclaw, Poland.} \affiliation{Laboratory for Solid State Physics, ETH Z\"urich,
CH-8093 Z\"{u}rich, Switzerland}

\author{M.~Chikovani}
\affiliation{Department of Physics, Tbilisi State University,
Chavchavadze 3, GE-0128 Tbilisi, Georgia}

\author{H.~Luetkens}
\affiliation{Laboratory for Muon Spin Spectroscopy, Paul Scherrer Institute, CH-5232
Villigen PSI, Switzerland}

\author{S.~Katrych}
\altaffiliation{Current address: Institut de Physique de la Mati\'{e}re Condens\'{e}e (ICMP), 
Ecole Polytechnique F\'{e}d\'{e}rale de Lausanne (EPFL), CH-1015 Lausanne, 
Switzerland.} \affiliation{Laboratory for Solid State Physics, ETH Z\"urich,
CH-8093 Z\"{u}rich, Switzerland}


\author{J.~Karpinski}
\altaffiliation{Current address: Institut de Physique de la Mati\'{e}re Condens\'{e}e (ICMP), 
Ecole Polytechnique F\'{e}d\'{e}rale de Lausanne (EPFL), CH-1015 Lausanne, 
Switzerland.}\affiliation{Laboratory for Solid State Physics, ETH Z\"urich,
CH-8093 Z\"{u}rich, Switzerland}

\author{H.~Keller}
\affiliation{Physik-Institut der Universit\"{a}t Z\"{u}rich,
Winterthurerstrasse 190, CH-8057 Z\"{u}rich, Switzerland}

\begin{abstract}
 
 The magnetic phase diagram of EuFe$_{2}$(As$_{1-x}$P$_{x}$)$_{2}$ 
was investigated by means of magnetization and muon-spin rotation (${\mu}$SR) studies
as a function of chemical (isovalent substitution of As by P) and hydrostatic pressure. 
The magnetic phase diagrams of the magnetic ordering of the Eu and Fe spins 
with respect to P content and hydrostatic pressure are determined and discussed.
The present investigations reveal that the magnetic coupling between the Eu and the Fe
sublattices strongly depends on chemical and hydrostatic pressure.
It is found that chemical and hydrostatic pressure have a similar
effect on the Eu and Fe magnetic order. 

\end{abstract}


\maketitle

\section{Introduction}

 The discovery of superconductivity in the iron-based
pnictides \cite{Kamihara08} provided a new class of compounds to the
high temperature superconductor (HTS) family. 
Ternary iron arsenide $A$Fe$_{2}$As$_{2}$
($A$ = Sr, Ca, Ba, Eu) \cite{Rotter} is one of the parent compounds with
ThCr$_{\rm 2}$Si$_{\rm 2}$-type structure. In analogy with $Ln$FeAsO
($Ln$ = La-Gd),\cite{Chen} $A$Fe$_{2}$As$_{2}$ undergoes a structural phase
transition from a tetragonal to an orthorombic phase, accompanied or followed
by a spin-density-wave (SDW) transition of the itinerant Fe moments.  
The superconducting (SC) state can be achieved either under pressure (chemical and hydrostatic) \cite{Torikachvili,Fukazawa,Terashima}
or by appropriate charge carrier doping of the parent compounds, \cite{RenLu08,Matsuishi08,Zhao08} both
accompanied by a suppression of the SDW state. 

  Here, we focus on EuFe$_{2}$As$_{2}$ which is a particularly interesting member of the ternary 
system $A$Fe$_{2}$As$_{2}$, since the $A$ site is occupied by a rare earth Eu$^{2+}$ $S$-state (orbital moment $L$ = 0) ion with a 4$f$$^{7}$ electronic configuration. Eu$^{2+}$ has a total
electron spin $S$ = 7/2, corresponding to a theoretical
effective magnetic moment of $\mu_{\rm eff}$ = 7.94 ${\mu}$$_{\rm B}$. In addition to the 
SDW ordering of the Fe moments at $T_{\rm SDW}$ ${\simeq}$ 190 K, an antiferromagnetic (AFM) 
order of the Eu$^{2+}$ spins 
at $T_{\rm AFM}$ ${\simeq}$ 19~K was reported by M\"{o}ssbauer spectroscopy \cite{Raffius93} and 
later confirmed by neutron diffraction.\cite{Xiao09}
Various reports on EuFe$_{2-x}$Co$_{x}$As$_{2}$ ($x$ = 0 and 0.1) suggest a 
strong coupling between the magnetism of the Eu$^{2+}$ ions and the conduction electrons,
which may affect or even destroy superconductivity.\cite{SJiang,Guguchia}
For example, in contrast to the
other '122' systems, where the substitution of Fe by Co leads to
superconductivity, \cite{Sefat,Jasper} the compounds containing Eu$^{2+}$ exhibit the onset of a superconducting
transition, but seem to be hindered to reach zero resistivity at ambient pressure. \cite{He08} 
Although Ni doping in BaFe$_{2}$As$_{2}$ leads to superconductivity up to 21 K,\cite{LiLuo08}
ferromagnetism rather than superconductivity was found
in EuFe$_{2}$As$_{2}$ by Ni doping.\cite{ZRen09}
On the other hand, in single crystals of P substituted EuFe$_{2}$(As$_{1-x}$P$_{x}$)$_{2}$
bulk superconductivity with superconducting transition temperature $T_{\rm c}$ ${\simeq}$ 28 K
was observed by resistivity, magnetization, and specific heat measurements.\cite{Jeevan} 
Isovalent P substitution on the As site in EuFe$_{2}$As$_{2}$ without introducing
holes or electrons, simulates a condition generally referred to as "chemical pressure".
Superconductivity coexisting with AFM Eu$^{2+}$ order
was only found in a very narrow range of P content $x$ 
(0.16 ${\leq}$ $x$ ${\leq}$ 0.22), where the SDW transition is suppressed.    
Superconductivity with a zero resistivity state was also observed for 
EuFe$_{2}$As$_{2}$ under applied pressure.\cite{Terashima,Matsubayashi} 
Similar to the case of P substitution, superconductivity covers only a 
narrow pressure range of 2.5-3.0 GPa.\\

 In this paper we report detailed magnetization and muon spin rotation (${\mu}$SR)
measurements in EuFe$_{2}$As$_{2-x}$P$_{x}$ as a function of the P content $x$.
One P substituted sample EuFe$_{2}$(As$_{0.88}$P$_{0.12}$)$_{2}$ was also studied  under applied
pressure $p$. The ${\mu}$SR technique is a powerful tool to study
the magnetic and superconducting properties of materials microscopically. It provides reliable
measurements of $T_{\rm c}$, $T_{\rm SDW}$, the magnetic ordering temperature of Eu$^{2+}$ spins
$T_{\rm Eu}$ and the ordered moment size as 
a function of both $x$ and $p$. Consequently, the 
phase diagrams with respect $x$ and $p$ are determined from these measurements.  
We compare the present results with previous high pressure studies on the parent compound
EuFe$_{2}$As$_{2}$ and discuss the combined results in terms of
the relation of $x$ and $p$.  
The paper is organized as follows: Experimental details are described in Sec.~II. 
The results of the magnetic susceptibility and the ${\mu}$SR experiments at ambient and
applied pressure are presented and discussed in Sec.~III(A) and III(B), respectively. 
In Sec.~IV the phase diagrams 
are presented. The conclusions follow in Sec.~V.

\section{EXPERIMENTAL DETAILS }
  In the present work the system EuFe$_{2}$(As$_{1-x}$P$_{x}$)$_{2}$ with $x$ = 0, 0.12, 0.2, and 1 
is investigated. Note that the sample with $x$ = 0 is single crystalline, and 
all the P substituted compounds are pollycrystalline.
The concentrations $x$ = 0.12, and 0.2 were studied due to their proximity  
to the SC phase reported in Ref.~19. 
A single crystal of EuFe$_{2}$As$_{2}$ was grown out of Sn flux.\cite{Guguchia2011} 
Polycrystalline samples were synthesized by solid-state reaction between EuAs, Fe$_{2}$As, and Fe$_{2}$P. 
EuAs was presynthesized by heating europium
grains and phosphorus powders very slowly to 1173 K followed by a tempering at this temperature for 36 h. 
Fe$_{2}$As was prepared by heating Fe and As powders
at 973 K for 10 h and at 1173 K for 15 h. Fe$_{2}$P was presynthesized by reacting iron
and phosphorus powders at 973 K for 24 h from stoichiometric 
amounts of the elements. All the starting materials
had a purity better than 99.9 ${\%}$. Powders of EuAs, Fe$_{2}$As and Fe$_{2}$P
were weighted according to the stoichiometric ratio, thoroughly ground and pressed into 
pellets in an argon-filled glove box. The pellets were then sealed in an evacuated
quartz tube, sintered at 1273~K for 36 h, and then cooled slowly to room temperature.
 
 Powder X-ray diffraction (XRD) studies of the EuFe$_{2}$(As$_{1-x}$P$_{x}$)$_{2}$ samples 
were carried out at room temperature with a 
STOE diffractometer (CuK$_{\alpha1}$ radiation, ${\lambda}$ = 1.5406 \AA) equipped with 
a mini-phase-sensitive detector and a Ge monochromator. The structural refinements
were done using the program FULLPROF.\cite{Rodriges}
The zero-field-cooled and field-cooled (ZFC and FC) magnetization measurements of the
EuFe$_{2}$(As$_{1-x}$P$_{x}$)$_{2}$ samples 
were performed with a commercial SQUID magnetometer ($Quantum$ $Design$ MPMS-XL).  
The samples with $x$ = 0.2 and 1 were studied only at ambient pressure.   
For $x$ = 0.12, the investigations were also carried out under applied pressures up to $p$ = 5.9 GPa
by using a diamond anvil cell (DAC) filled with Daphne oil which served as a
pressure-transmitting medium. The pressure at low temperatures was determined 
by the pressure dependence of the SC transition temperature of Pb.

 Zero-field (ZF) ${\mu}$SR experiments were performed at the ${\mu}$E1 and ${\pi}$M3 
beamlines of the Paul Scherrer Institute (Villigen, Switzerland). The general purpose instrument
(GPS) was used to study the system EuFe$_{2}$(As$_{1-x}$P$_{x}$)$_{2}$ ($x$ = 0, 0.12, 0.2, and 1)
at ambient pressure. The samples were mounted inside of a gas-flow
$^{4}$He cryostat on a sample holder with a standard veto setup providing 
a low-background ${\mu}$SR signal. In addition, the sample with $x$ = 0.12 was studied under pressure
using the GPD instrument. Pressures up to 2.0 GPa were generated in a double wall
piston-cylinder type of cell made of MP35N \cite{Andreica} material especially designed to perform ${\mu}$SR
experiments under pressure. As a pressure transmitting medium Daphne oil was used. 
The pressure was measured by tracking the SC transition of a very small indium plate by AC susceptibility.
The ${\mu}$SR time spectra were analyzed using the free software package MUSRFIT.\cite{Suter}

\section{RESULTS AND DISCUSSION}
\subsection{Crystal structure and magnetic properties of EuFe$_{2}$(As$_{1-x}$P$_{x}$)$_{2}$}
\subsubsection{X-ray powder diffraction}
\begin{figure}[b!]
\includegraphics[width=1.0\linewidth]{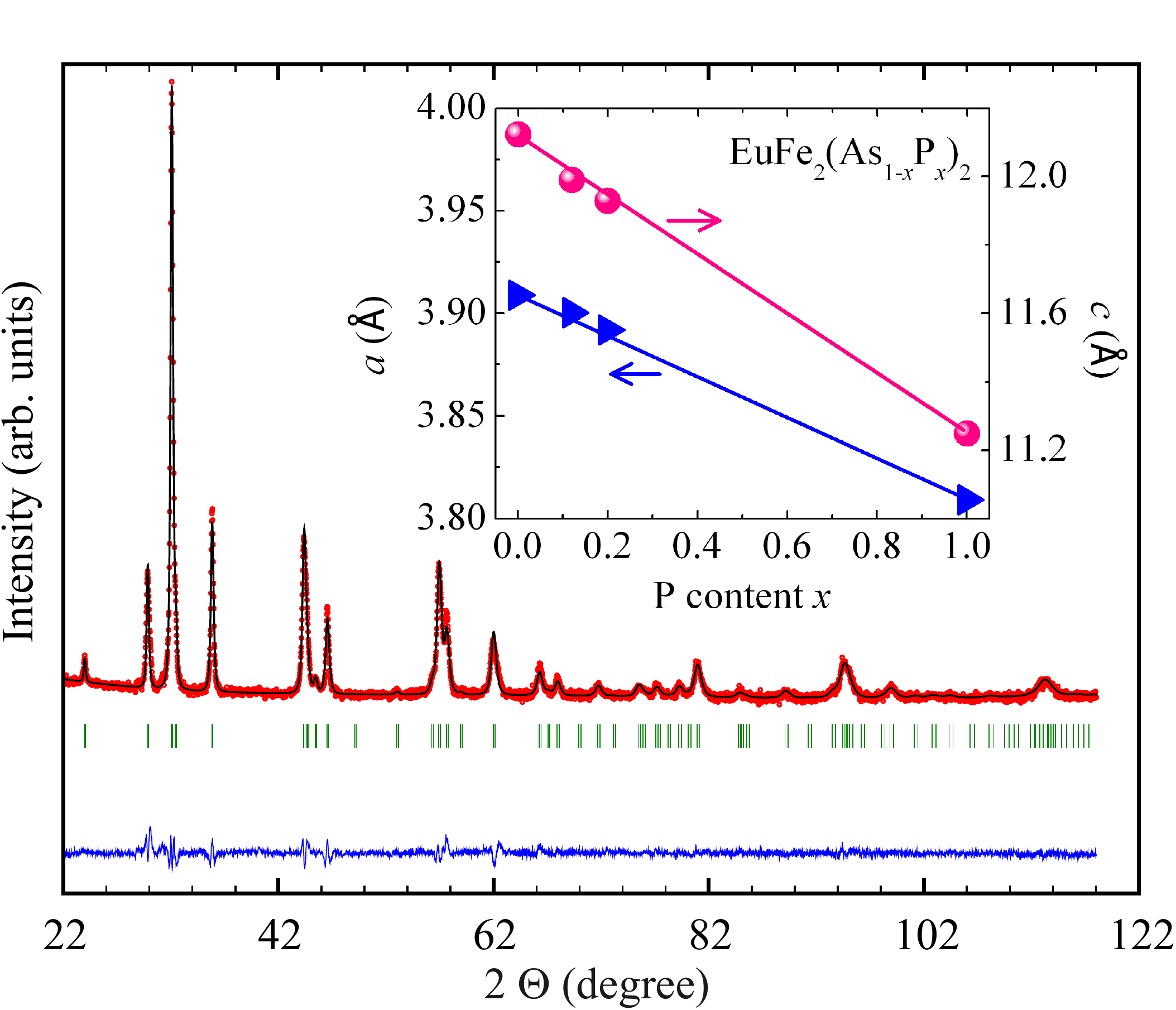}
\vspace{-0.5cm}
\caption{ (Color online) X-ray powder diffraction pattern at room temperature for the sample 
EuFe$_{2}$(As$_{0.88}$P$_{0.12}$)$_{2}$. The solid line represents a Ritveld refinement profile. The residuals
are plotted at the bottom of the figure. In the inset refined  lattice parameters are plotted 
as a function of P content $x$.}
\label{fig1}
\end{figure}
The crystal structure for all EuFe$_{2}$(As$_{1-x}$P$_{x}$)$_{2}$ samples 
at room temperature was refined with the tetragonal ThCr$_{2}$Si$_{2}$ structure. 
An example of the refinement profile for EuFe$_{2}$(As$_{0.88}$P$_{0.12}$)$_{2}$ is
shown in Fig.~1. No obvious secondary phase can
be detected. The weighted pattern factor and goodness
of fit are $R{\rm wp}$ $\sim$ 11.2 ${\%}$  and S $\sim$ 1.6, respectively, indicating 
a fairly good refinement. In addition, the refined
occupancies are close to the nominal values. 
The lattice constants for the tetragonal unit cell based upon the Rietveld refinements are 
$a$ = 3.9095(2) \AA~ and $c$ = 11.979(1) \AA~ for $x$ = 0.12, $a$ = 3.9006(2) \AA~ and $c$ = 11.9312(1) \AA~ for $x$ = 0.2,
$a$ = 3.8152(2) \AA~ and $c$ = 11.2401(1) \AA~ for $x$ = 1. The values for $x$ = 1 are in agreement with the literature values 
[$a$ = 3.8178(1) \AA~ and $c$ = 11.2372(3) \AA].\cite{GCao}
The lattice constants $a$ and $c$ as a function of $x$ are plotted in the inset of Fig.~1. A decrease of both $a$ and $c$ 
with increasing $x$ is observed. The decrease of the lattice constant $c$ as a result of P substitution 
implies an increase of the coupling between the Eu and the Fe$_{2}($As$_{1-x}$P$_{x}$)$_{2}$ layers.   
This might also be important for the evolution of the magnetic order in the Eu-sublattice,
since the Ruderman-Kittel-Kasuya-Yosida (RKKY) coupling strongly depends on the distance
between the magnetic ions.\cite{ZRen09,Guguchia,GCao}

\subsubsection{Magnetization measurements}
 The temperature dependence of the zero-field-cooled (ZFC) and field-cooled (FC)  
magnetic susceptibility ${\chi}$ = $M$/$H$ 
for EuFe$_{2}$(As$_{1-x}$P$_{x}$)$_{2}$ ($x$ = 0.12, 0.2, and 1) 
in a magnetic field of ${\mu}_{\rm 0}$$H$ = 2 mT 
is shown in Fig.~2.
The results for $x$ = 0 were already discussed in detail in our previous work,\cite{Guguchia2011}
and hence, are not shown here.
The magnetic susceptibility at high temperatures ($i.e.,$ far above  the
magnetic ordering temperature of the Eu$^{2+}$ moments $T_{\rm Eu}$)
is well described by the Curie-Weiss law:
\begin{equation}
\chi_{\rm}(T)=\frac{C}{T-\theta_{\rm CW}} \label{eq1}.
\end{equation}
 Here, ${\textit C}$ denotes the Curie constant and ${\theta}$$_{\rm CW}$ the paramagnetic Curie-Weiss temperature.
An analysis of the data in Fig.~2 with Eq.~(1) yields: ${\theta}$$_{\rm CW}$ = 16.74(8) K,
${\mu}$$_{\rm eff}$ ${\simeq}$ 8.1 ${\mu}$$_{\rm B}$ for $x$ = 0.12, 
${\theta}$$_{\rm CW}$ = 18.14(7) K, ${\mu}$$_{\rm eff}$ ${\simeq}$ 8.2 ${\mu}$$_{\rm B}$ for $x$ = 0.2, and 
${\theta}$$_{\rm CW}$ = 29.35(9) K, ${\mu}$$_{\rm eff}$ ${\simeq}$ 8.3 ${\mu}$$_{\rm B}$ for $x$ = 1.   
The obtained values of ${\mu}$$_{\rm eff}$ are close to the theoretical value of 
a free Eu$^{2+}$ ion (${\mu}$$_{\rm Eu^{2+}}$ = 7.94~${\mu}$$_{\rm B}$).
\begin{figure}[b!]
\includegraphics[width=1.0\linewidth]{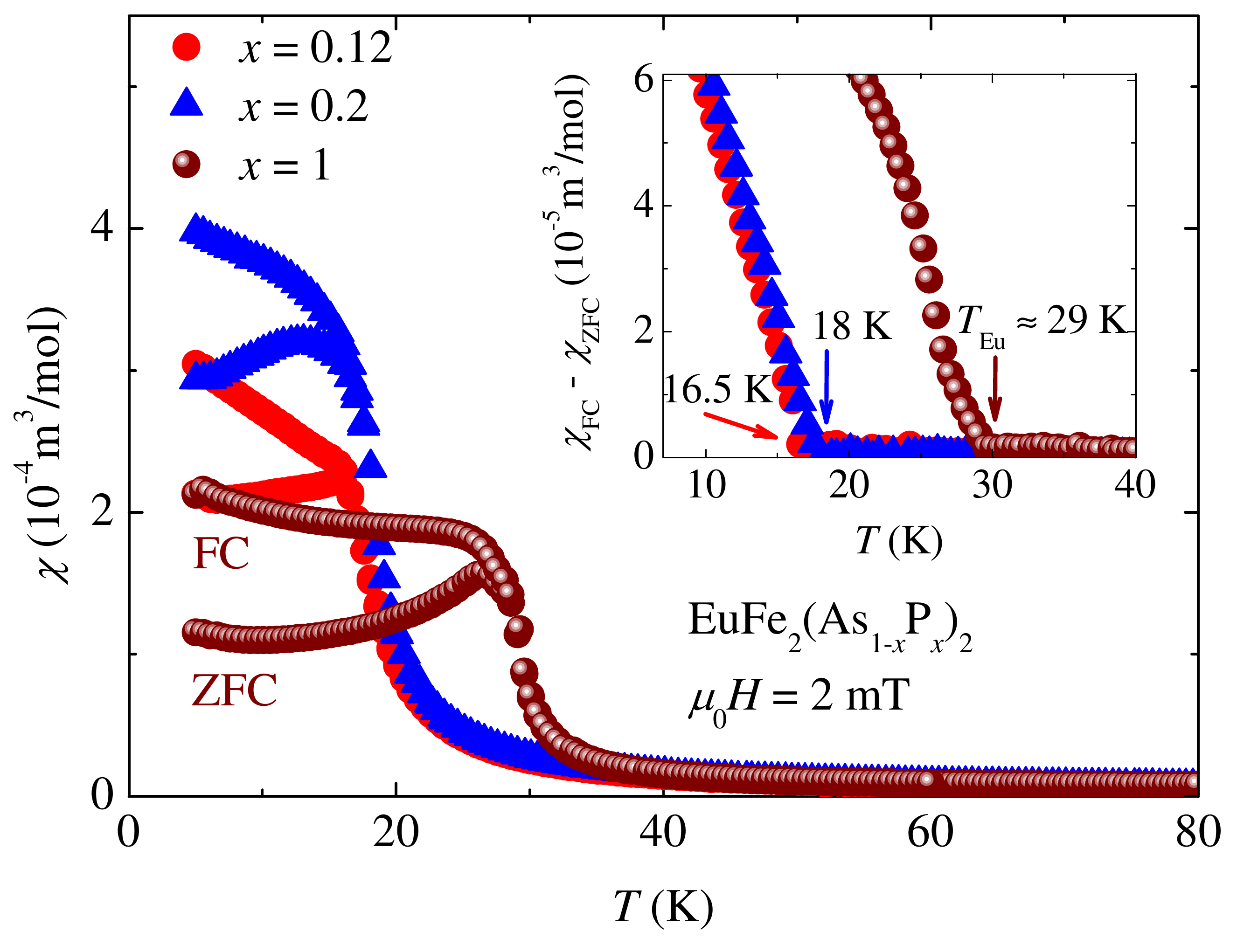}
\caption{ (Color online) Temperature dependence of the ZFC and FC magnetic susceptibility 
for the samples EuFe$_{2}$(As$_{1-x}$P$_{x}$)$_{2}$ ($x$ = 0.12, 0.2, 1) 
measured in a magnetic field of ${\mu}_{\rm 0}$$H$ = 2 mT. The inset illustrates
the temperature dependence of the difference of both susceptibilities 
(${\chi}$$_{\rm FC}$-${\chi}$$_{\rm ZFC}$).
The arrows mark the magnetic ordering temperatures $T_{\rm Eu}$ of the Eu$^{2+}$ moments.}
\label{fig1}
\end{figure}

 As shown in Fig.~2 for all the P substituted samples an obvious deviation between ${\chi}_{\rm ZFC}$
and ${\chi}_{\rm FC}$ is seen at low temperatures. This is not the case for $x$ = 0,\cite{ZRen09,Guguchia2011}
for which AFM order of Eu$^{2+}$ with the moments pointing along the $a$ axis was reported. 
This result is consistent with previous magnetizaton studies,\cite{Sina} suggesting that 
the ground state of the coupled Eu$^{2+}$ spins is a canted AFM state (C-AFM state) (i.e., AFM with the net ferromagnetic (FM) component along the
$c$-axis) in EuFe$_{2}$(As$_{1-x}$P$_{x}$)$_{2}$ ($x$ = 0.12, 0.2) and a FM state in EuFe$_{2}$P$_{2}$.  
Recently, neutron diffraction measurements were also performed on EuFe$_{2}$P$_{2}$ and an
almost axial FM structure of the Eu$^{2+}$ spins was established.\cite{Ryan} The C-AFM and FM structure of the Eu-sublattice
in EuFe$_{2}$(As$_{1-x}$P$_{x}$)$_{2}$ ($x$ = 0.12, 0.2, 1) sharply contrasts with the planar 
antiferromagnetism seen in the parent compound EuFe$_{2}$As$_{2}$, suggesting a delicate interplay between
the Eu 4$f$ and the Fe 3$d$ electrons. 
It was concluded from different experiments \cite{SJiang,Guguchia}
that there is a strong coupling between the localized Eu$^{2+}$ 
spins and the conduction electrons of the two-dimensional (2D) Fe$_{2}$As$_{2}$ layers in 
EuFe$_{2}$As$_{2}$. This revealed that the magnetic exchange interaction between 
the localized Eu 4$f$ moments is mediated by the itinerant Fe 3$d$ electrons.
However, the interaction of the Eu moments with the 
magnetic moments of the Fe sublattice (band magnetism) cannot be neglected.
Only a combination of both interactions can further elucidate 
the C-AFM ground state observed in  EuFe$_{2}$(As$_{1-x}$P$_{x}$)$_{2}$ ($x$ = 0.12 and 0.2).
Note that a C-AFM ground state was also found in the related 
compound EuFe$_{1.8}$Co$_{0.2}$As$_{2}$.\cite{Guguchia2011}

 The magnetic ordering temperature $T_{\rm Eu}$  
of the Eu$^{2+}$ moments was determined by the temperature at which the difference between
${\chi}_{\rm ZFC}$ and ${\chi}_{\rm FC}$ sets in (see inset of Fig.~2). It was found to be 
$T_{\rm Eu}$ ${\simeq}$ 16.5 K, 18 K, and 29 K for $x$ = 0.12, $x$ = 0.2, and $x$ = 1, respectively.
The value of $T_{\rm Eu}$ for $x$ = 0.12 is slightly reduced compared to $T_{\rm Eu}$ ${\simeq}$ 19~K  for the parent 
compound $x$ = 0. However, on further increasing the P concentration $T_{\rm Eu}$ increases and reaches a maximum 
for $x$ = 1. The value of $T_{\rm Eu}$ for $x$ = 1 is in agreement with those reported in literature.\cite{Jeevan,GCao,Ryan}

\subsubsection{Zero-field ${\mu}$SR measurements}
\begin{figure*}[ht!]
\centering
\includegraphics[width=0.8\linewidth]{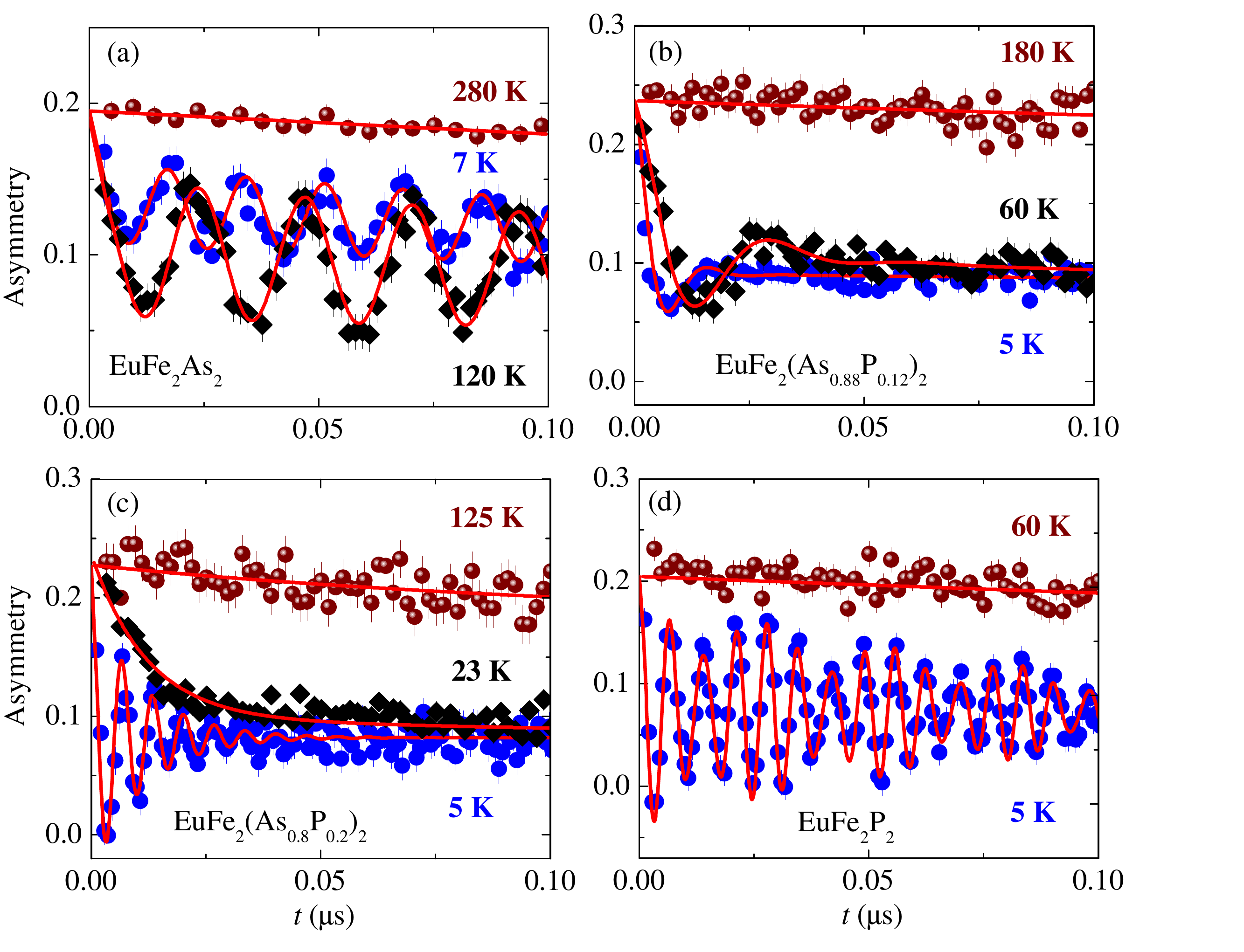}
\vspace{0cm}
\caption{ (Color online) ZF ${\mu}$SR spectra for EuFe$_{2}$(As$_{1-x}$P$_{x}$)$_{2}$ 
($x$ = 0, 0.12, 0.2, 1) recorded for three different temperatures: $T$ ${\textless}$ $T_{\rm Eu}$ (circles), 
$T_{\rm Eu}$ ${\textless}$ $T$ ${\textless}$ $T_{\rm SDW}$ (diamonds), 
and $T$ ${\textgreater}$ $T_{\rm SDW}$ (spheres). 
The solid lines represent fits to the data by means of Eq.~(2).}
\label{fig3}
\end{figure*}

 In a ${\mu}$SR experiment nearly 100 ${\%}$ spin-polarized muons ${\mu}$$^{+}$
are implanted into the sample one at a time. The positively
charged ${\mu}$$^{+}$ thermalize at interstitial lattice sites, where they
act as magnetic microprobes. In a magnetic material the 
muon spin precesses in the local magnetic field $B_{\rm \mu}$ at the
muon site with the Larmor frequency ${\nu}_{\rm \mu}$ = $\gamma_{\rm \mu}$/(2${\pi})$$B_{\rm \mu}$ (muon
gyromagnetic ratio $\gamma_{\rm \mu}$/(2${\pi}$) = 135.5 MHz T$^{-1}$). 
ZF ${\mu}$SR is a very powerfull tool 
to investigate microscopic magnetic properties of solids without applying an external magnetic field.\\ 
 
 ZF ${\mu}$SR time spectra for the single crystal of EuFe$_{2}$As$_{2}$ and
for the polycrystalline samples EuFe$_{2}$(As$_{1-x}$P$_{x}$)$_{2}$ 
are shown in Fig.~3, recorded for three different temperatures: 
$T$ ${\textless}$ $T_{\rm Eu}$, $T_{\rm Eu}$ ${\textless}$ $T$ ${\textless}$ $T_{\rm SDW}$, 
and $T$ ${\textgreater}$ $T_{\rm SDW}$. 
For EuFe$_{2}$As$_{2}$ the ZF ${\mu}$SR measurements were performed with the initial
muon spin polarization tilted by approximately 45$^{\circ}$ away from the crystallographic 
$c$-axis. At high temperatures (see Fig.~3), no muon spin precession and only a very 
weak depolarization of the ${\mu}$SR signal is observed. This weak
depolarization and its Gaussian functional form are typical for a paramagnetic material 
and reflect the occurrence of a small Gaussian-Kubo-Toyabe depolarization, originating from the
interaction of the muon spin with randomly oriented nuclear
magnetic moments. At temperatures below $T_{\rm Eu}$ a well-defined spontaneous muon spin 
precession is observed in all compounds, indicating 
long-range magnetic order of the Eu$^{2+}$ moments in the investigated
compounds. For $x$ = 0 and 0.12 above $T_{\rm Eu}$ ${\simeq}$ 20.5 K and 16.5 K, respectively,
muon spin precession with a lower frequency is observed which is caused by the long-range 
SDW order of the Fe moments.  
However, for $x = 0.2$, instead of the oscillatory behavior seen in the SDW state
for $x$ = 0 and 0.12, a fast decaying signal is observed (see Fig.~3c). 
The reason for this strongly decaying ${\mu}$SR signal will be discussed below.
For $x$ = 1 only the magnetic ordering of the Eu moments is seen in the ${\mu}$SR spectra (Fig.~3d).
Note that for $x$ = 0, 0.12 and 0.2 only one ${\mu}$SR frequency  
is visible. However, for $x$ = 1 two distinct precession frequencies occur in the
${\mu}$SR spectra, corresponding to the local magnetic fields $B^{1}_{\rm \mu,Eu}$ ${\simeq}$ 1.08~T
(${\simeq}$ 70${\%}$ of the signal) and $B^{2}_{\rm \mu,Eu}$ ${\simeq}$ 1.37 T (${\simeq}$ 30${\%}$
of the signal). This indicates that
two magnetically inequivalent muon stopping sites are
present in EuFe$_{2}$P$_{2}$. 
  
 The ZF ${\mu}$SR data were analyzed using the following
functional form:
\begin{equation}
A(t)=\sum_{i=1}^{2} {A_0^{i}\Bigg[{\alpha}_{i}e^{-\lambda^{i}_{T}t}\cos(\gamma_{\mu}B^{i}_{\mu}t+\varphi)}+{\beta}_{i}e^{-\lambda^{i}_{L}t}\Bigg]. 
\label{eq1}
\end{equation}
${\alpha}_{\rm i}$ and ${\beta}{\rm i}$ = 1 - ${\alpha}_{\rm i}$ (i = 1 for $x$ = 0, 0.12, 0.2, and i = 1, 2 for $x$ = 1) 
are the fractions of the oscillating and nonoscillating ${\mu}$SR signal.
For the single crystal ($x$ = 0) one finds ${\alpha}_{\rm 1}$ = 0.73(2) and ${\beta}_{\rm 1}$ = 0.27(3). 
However, for the polycrystalline samples ${\alpha}_{\rm i}$ = 2/3 
and ${\beta}{\rm i}$ = 1/3.
The 2/3 oscillating and the 1/3 nonoscillating ${\mu}$SR
signal fractions originate from the spatial averaging in 
powder samples where only 2/3 of the magnetic field components are
perpendicular to the muon spin and cause muon spin precession. 
$A_{\rm 0}$ denotes the initial asymmetry, 
and ${\varphi}$ is
the initial phase of the muon-spin ensemble. $B^{i}_{\rm \mu}$ represents the internal magnetic field
at the muon site, and the depolarization rates ${\lambda}^{i}_{\rm T}$ and ${\lambda}^{i}_{\rm L}$ characterize the 
damping of the oscillating and nonoscillating part of the ${\mu}$SR signal, respectively.
The transversal relaxation rate ${\lambda}^{i}_{\rm T}$ is a measure of the width of the static magnetic field
distribution at the muon site, and also reflects dynamical effects (spin fluctuations). 
The longitudinal relaxation rate ${\lambda}^{i}_{\rm L}$ is determined by dynamic magnetic fluctuations only.\cite{maeter} 
\begin{figure}[t!]
\includegraphics[width=1.0\linewidth]{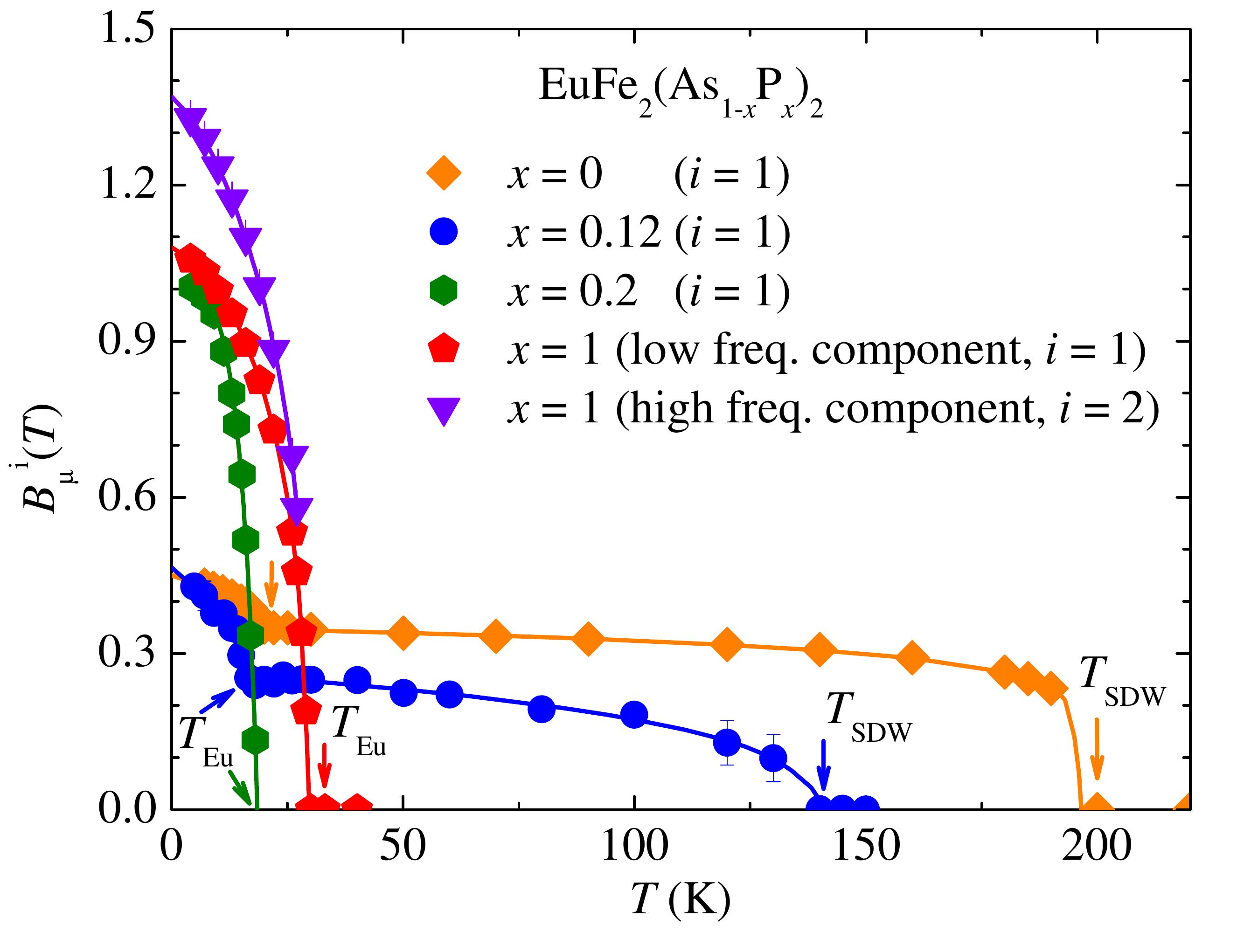}
\vspace{-0.8cm}
\caption{ (Color online) The temperature dependence of the internal magnetic field
$B^i_{\rm \mu}$ for the samples EuFe$_{2}$(As$_{1-x}$P$_{x}$)$_{2}$ ($x$ = 0, 0.12, 0.2, and 1).
The solid lines represent fits to the data by means of Eq.~(3). The arrows mark the transition temperatures
for the SDW ($T_{\rm SDW}$) and the Eu magnetic order ($T_{\rm Eu}$).}
\label{fig1}
\end{figure}
\begin{figure}[b!]
\includegraphics[width=1.22\linewidth]{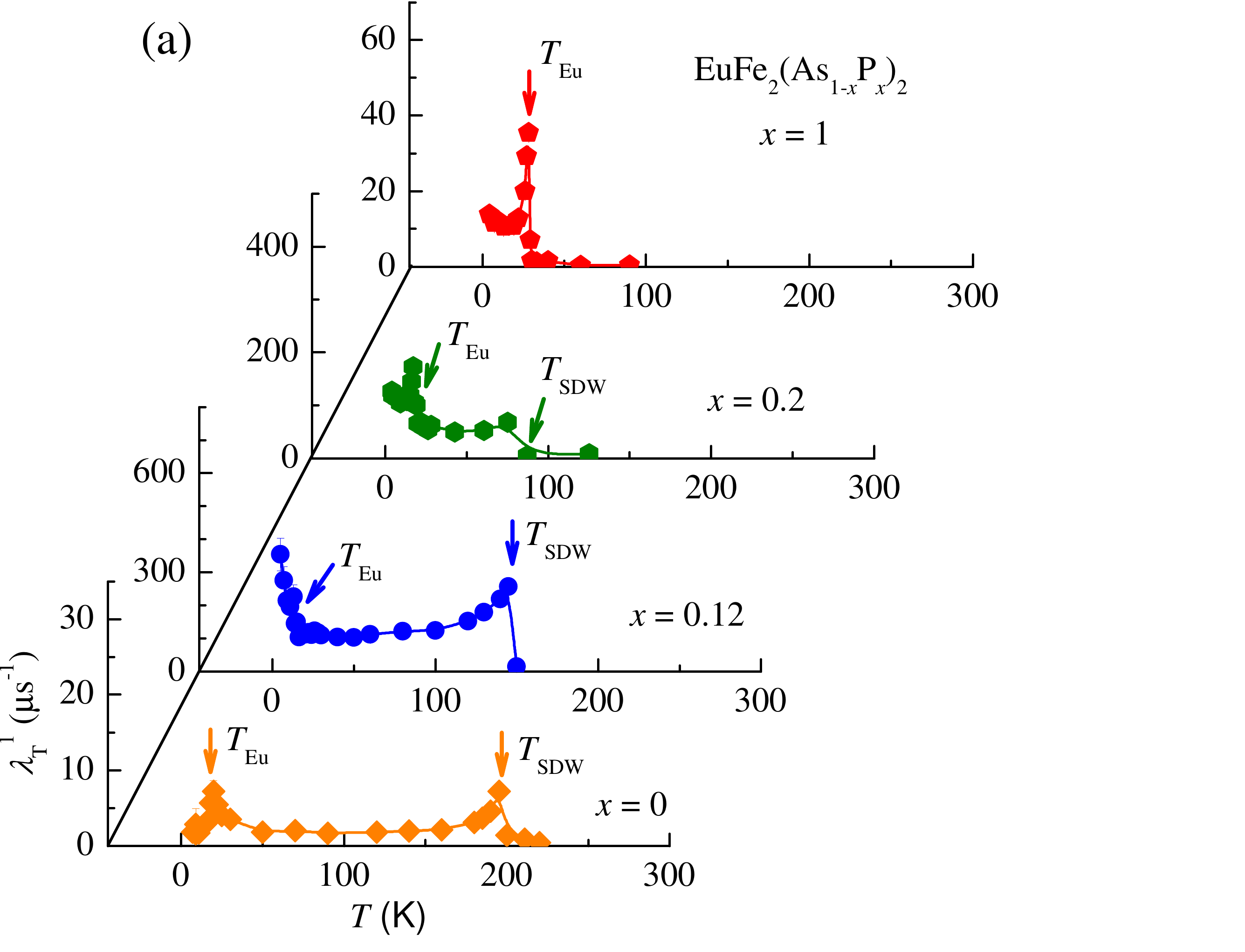}
\includegraphics[width=1.22\linewidth]{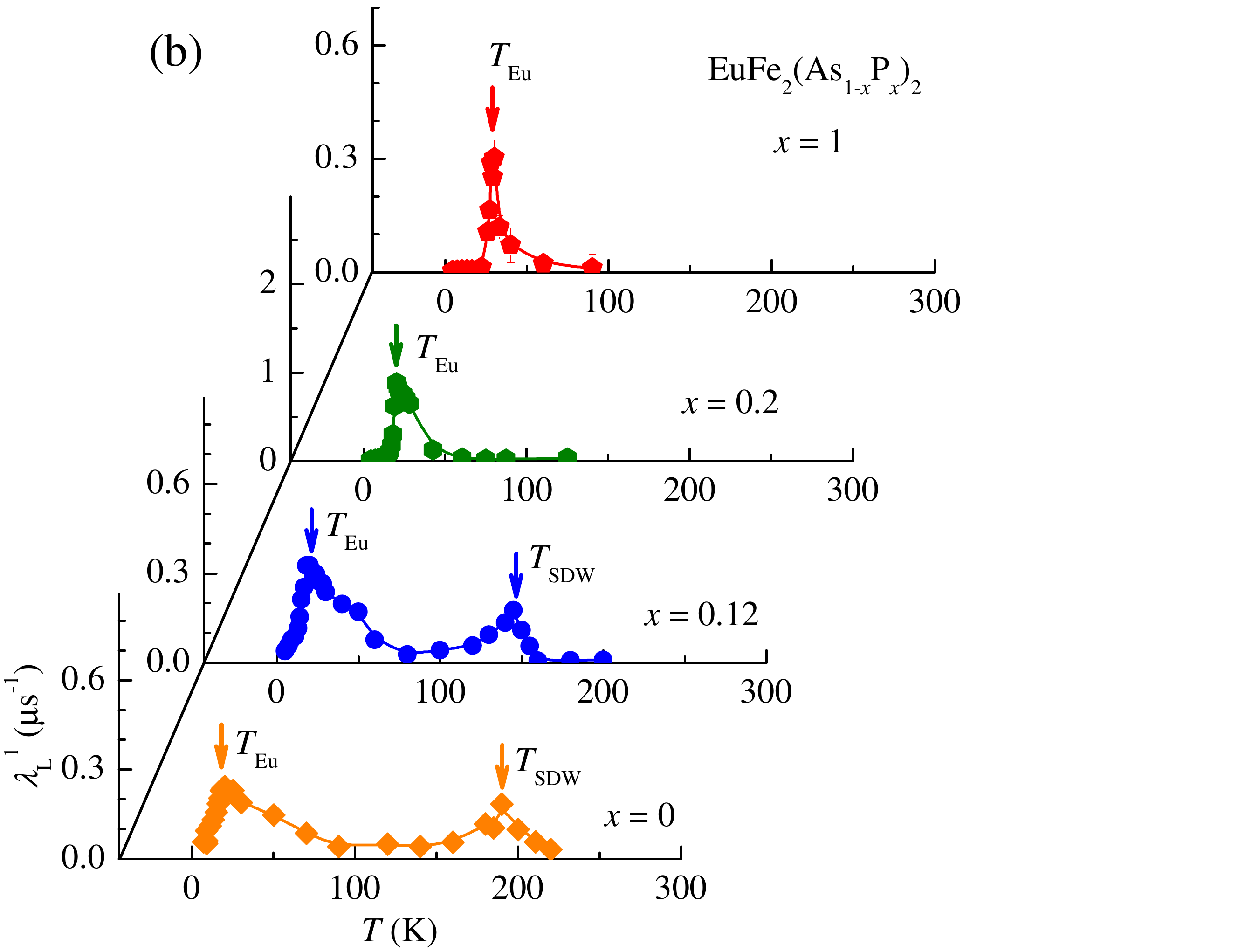}
\vspace{-0.3cm}
\caption{ (Color online) (a) Transverse relaxation rate ${\lambda}^{1}_{\rm T}$~($T$) 
for the samples EuFe$_{2}$(As$_{1-x}$P$_{x}$)$_{2}$ with $x$ = 0, 0.12, 0.2, and 1. 
Lines are guides to the eye. (b) Longitudinal relaxation rate ${\lambda}^{1}_{\rm L}$ ($T$)
for EuFe$_{2}$(As$_{1-x}$P$_{x}$)$_{2}$ ($x$ = 0, 0.12, 0.2, 1). The arrows mark the transition temperatures
for the high-temperature SDW ($T_{\rm SDW}$) and the low-temperature Eu order ($T_{\rm Eu}$).}
\label{fig1}
\end{figure}
 The temperature dependence of the internal magnetic field $B^i_{\rm \mu}$  
for EuFe$_{2}$(As$_{1-x}$P$_{x}$)$_{2}$ is shown in Fig.~4.
$B^i_{\rm \mu}$ is proportional to the magnitude of the ordered moment and thus to the
magnetic order parameter. The second component ($i$ = 2) in the ${\mu}$SR signal was observed
only for $x$=1, and hence, we will discuss the $x$-dependence of the relevant physical parameters
related to the first component ($i$ = 1) only.  
For $x$ = 0 a sharp step like increase of $B^1_{\rm \mu}$ is observed below ${\simeq}$ 195 K,
which reflects the appearance of the SDW ordering of the Fe moments.
The value of $T_{\rm SDW}$ is in good agreement with  
$T_{\rm SDW}$ ${\simeq}$ 190 K obtained from neutron diffraction.\cite{Xiao09}
A sharp increase of $B^1_{\rm \mu}$ is an indication for a first order transition.
A first order transition due to SDW formation was also observed in the related compound SrFe$_{2}$As$_{2}$.\cite{Jesche} 
Upon lowering the temperature $B^1_{\rm \mu}$ first tends to saturate,
but increases again when the magnetic order of the Eu$^{2+}$ moments 
occurs at $T_{\rm Eu}$. 
\begin{table*}[t!]
\caption{Summary of the parameters obtained for the polycrystalline samples of 
EuFe$_{2}$(As$_{1-x}$P$_{x}$)$_{2}$ ($x$ = 0, 0.12, 0.2, 1) by means of magnetization and ${\mu}$SR experiments.
$T_{\rm Eu}^{\chi}$ and $T_{\rm Eu}^{{\mu}SR}$ are the magnetic ordering temperatures of the Eu moments determined by
susceptibility and ${\mu}$SR measurements. $T_{\rm SDW}$ denotes the SDW ordering temperature of the Fe moments
determined from ${\mu}$SR experiments. $B^i_{\rm \mu,Eu}$(0) and $B^i_{\rm \mu,SDW}$(0) ($i$ = 1, 2) represent the zero-temperature values of
the internal magnetic fields at the muon site probed in the Eu and the SDW ordered state.}
\vspace{0.3cm}
\begin{tabular}{lcccccccc}
\hline
\hline
& $x$  &$T_{\rm Eu}^{\chi}$ & $T_{\rm Eu}^{{\mu}SR}$ & $T_{\rm SDW}$ &  $B^{1}_{\rm \mu,Eu}$ (0) & $B^{2}_{\rm \mu,Eu}$ (0) & $B^1_{\rm \mu,SDW}$ (0) & $B^2_{\rm \mu,SDW}$ (0)\\
&       &  (K)  & (K) & (K) & (T) & (T) & (T) & (T) \\
\hline 
   &  0                           & 19.5(6)     &  20.5(5)   & 195(3)   & 0.45(1) & - & 0.35(1) & -\\    
   & 0.12                        & 16.5(5) & 16.7(6)  & 140(5) & 0.476(12) & - & 0.258(10) & -\\ 
   & 0.2                         & 17.9(5) & 18.4(2)  & 85 & 0.997(12) & - & 0 & -\\
   &  1                           & 29.5(5) & 29.3(4)  & 0 & 1.08(1) & 1.37(2) & 0 & 0\\                   \hline
\hline         
\end{tabular}
\label{table1}
\end{table*}
 To describe the temperature dependence of $B^i_{\rm \mu}$ we assumed the following phenomenological function:
\begin{equation}
\begin{split}
\vec{B}^i_{\mu}(T)=\vec{B}^i_{\mu,Eu}(0)\Bigg[1-\Bigg(\frac{T}{T_{Eu}}\Bigg)^{\gamma_1} \Bigg]^{\delta_1}\\
+\vec{B}^i_{\mu,SDW}(0)\Bigg[1-\Bigg(\frac{T}{T_{SDW}}\Bigg)^{\gamma_2} \Bigg]^{\delta_2},\\
\end{split}
\end{equation}
where $B^i_{\rm \mu,Eu}$(0) and $B^i_{\rm \mu,SDW}$(0) represent the zero-temperature values
of the internal magnetic field probed by the muons in the Eu and in the SDW ordered states, respectively.   
${\gamma}$ and ${\delta}$ are empirical exponents. 
As indicated by the solid lines in Fig.~4 the function in Eq.~(3) describes the data reasonably well, yielding
the parameters given in Table.~I. Note that with increasing $x$ the values of $T_{\rm SDW}$ 
and $B^1_{\rm \mu,SDW}$(0) decrease, and for $x$ = 0.2 and $x$ = 1 no long-range SDW order of the Fe moments is 
observed. On the other hand, $T_{\rm Eu}$ decreases with increasing $x$, reaches minimum at $x$ = 0.12 and 
then increases again, in agreement with the above susceptibility measurements.
In addition, $B^1_{\rm \mu,Eu}$(0) significantly increases with $x$ above $x$=0.12.
Considering the magnetization results, the increase of $B^1_{\rm \mu,Eu}$(0) may be ascribed
to the appearance/growth of the ferromagnetic component as a result of P substitution. 
However, without microscopic modeling (i.e., calculation of the ${\mu}$ stopping site and the dipolar fields 
at the ${\mu}$ site) it is not possible to conclude how a change of the magnetic structure with 
P substitution would affect the internal field at the muon site.  

 The temperature dependences of the transverse and longitudinal depolarization rates 
${\lambda}^{1}_{\rm T}$ and ${\lambda}^{1}_{\rm L}$ are 
presented in Figs.~5a and 5b, respectively. Note that ${\lambda}^{1}_{\rm T}$ is much smaller 
for the end members $x$ = 0, 1 of the investigated system than for the 
mixed compounds $x$ = 0.12, 0.2. 
As shown in Fig.~5a, for $x$ = 0, 0.12, and 0.2 the onset of the Fe
magnetic order is accompanied by an increase of ${\lambda}^{1}_{\rm T}$
that decreases with decreasing temperature. 
Upon reaching the magnetic ordering temperature of Eu
${\lambda}^{1}_{\rm T}$ shows another maximum. For $x$ = 1, the strong increase of ${\lambda}^{1}_{\rm T}$
around $T_{\rm Eu}$ is only due the Eu order. No SDW transition is observed at higher temperatures. 
The magnetic ordering temperatures of Eu ($T_{\rm Eu}$) and Fe ($T_{\rm SDW}$) are also clearly
visible in the longitudinal relaxation rate ${\lambda}^{1}_{\rm L}$, which also shows
a clear anomaly at $T_{\rm Eu}$ and $T_{\rm SDW}$ (see Fig.~5b). 
As mentioned above, for the sample with $x$ = 0.2 (see Fig.~3c, diamonds) only
a fast depolarization of the implanted muons is observed above $T_{\rm Eu}$, but no coherent 
precession signal. The fast depolarization of the ${\mu}$SR signal could be either due to a wide
distribution of static fields, and/or to strongly fluctuating 
magnetic moments. To discriminate between these two possibilities we compare the 
values of ${\lambda}^{1}_{\rm T}$ and ${\lambda}^{1}_{\rm L}$. 
Note that for $x$ = 0.2, and $T$ ${\textless}$ 85 K ${\lambda}^{1}_{\rm T}$ is 
very large (${\simeq}$ 50 MHz) while ${\lambda}^{1}_{\rm L}$ is small (${\simeq}$ 0.05 MHz).
${\lambda}^{1}_{\rm T}$ consists of a static as well as of a dynamic contribution, while 
${\lambda}^{1}_{\rm L}$ contains only a dynamic contribution. Since in our case 
${\lambda}^{1}_{\rm T}$ ${\gg}$ ${\lambda}^{1}_{\rm L}$,
the static contribution dominates ${\lambda}^{1}_{\rm T}$, and 
the fast depolarization of the ${\mu}$SR signal observed for $x$ = 0.2 is due to the 
(quasi-)static disordered SDW phase with $T_{\rm SDW}$ ${\simeq}$ 85 K. 
The important parameters for all samples extracted from the magnetization and the ${\mu}$SR experiments are summarized in
Table~I.\\
  Very recently, bulk superconductivity with $T_{\rm c}$ ${\simeq}$ 28 K
was reported in single crystals of P substituted EuFe$_{2}$(As$_{1-x}$P$_{x}$)$_{2}$, \cite{Jeevan} 
based on resistivity, magnetization, and specific heat measurements. 
However, superconductivity coexisting with AFM Eu$^{2+}$ order
was only found in a very narrow $x$ range 
(0.16 ${\leq}$ $x$ ${\leq}$ 0.22), where the SDW transition is suppressed.
In the present study no indication of superconductivity was seen for $x$ = 0.2 from
magnetization measurements.
This might be due to the fact that in our sample ($x$ = 0.2) the SDW state is not completely
suppressed as supported by the ${\mu}$SR measurements.
In addition to chemical pressure, the physical properties of 
EuFe$_{2}$As$_{2}$ can be also tuned by the application of hydrostatic pressure.\cite{Terashima,Matsubayashi}
Previous reports of high pressure experiments on EuFe$_{2}$As$_{2}$
revealed pressure-induced superconductivity in a narrow pressure range of 2.5-3.0 GPa,\cite{Terashima,Matsubayashi}
accompanied by a suppression of the SDW state of the Fe moments.
Since pressure experiments on EuFe$_{2}$As$_{2}$ were already reported by various groups,\cite{Terashima,Matsubayashi}
we decided to study pressure effects in the P substituted sample EuFe$_{2}$(As$_{0.88}$P$_{0.12}$)$_{2}$. 
The sample with $x$ = 0.12 was chosen for the following reasons:
1) according to the SC phase diagram reported \cite{Jeevan} for  
EuFe$_{2}$As$_{2}$ as a function of chemical pressure (P content $x$) 
the sample with $x$ = 0.12 is close to the value of $x$ at which superconductivity
appears. By applying hydrostatic pressure, the SC phase might be reachable.
2) based on previous reports,\cite{Terashima,Matsubayashi} superconductivity was found in the vicinity
of the pressure value where the SDW state is suppressed.\\ 
   In the following sections the results of the magnetization and the ${\mu}$SR experiments
performed on EuFe$_{2}$(As$_{0.88}$P$_{0.12}$)$_{2}$ under hydrostatic
pressures are presented.

\subsection{Hydrostatic pressure effect on EuFe$_{2}$(As$_{0.88}$P$_{0.12}$)$_{2}$}

\subsubsection{High pressure magnetization measurements} 
\begin{figure}[t!]
\includegraphics[width=0.9\linewidth]{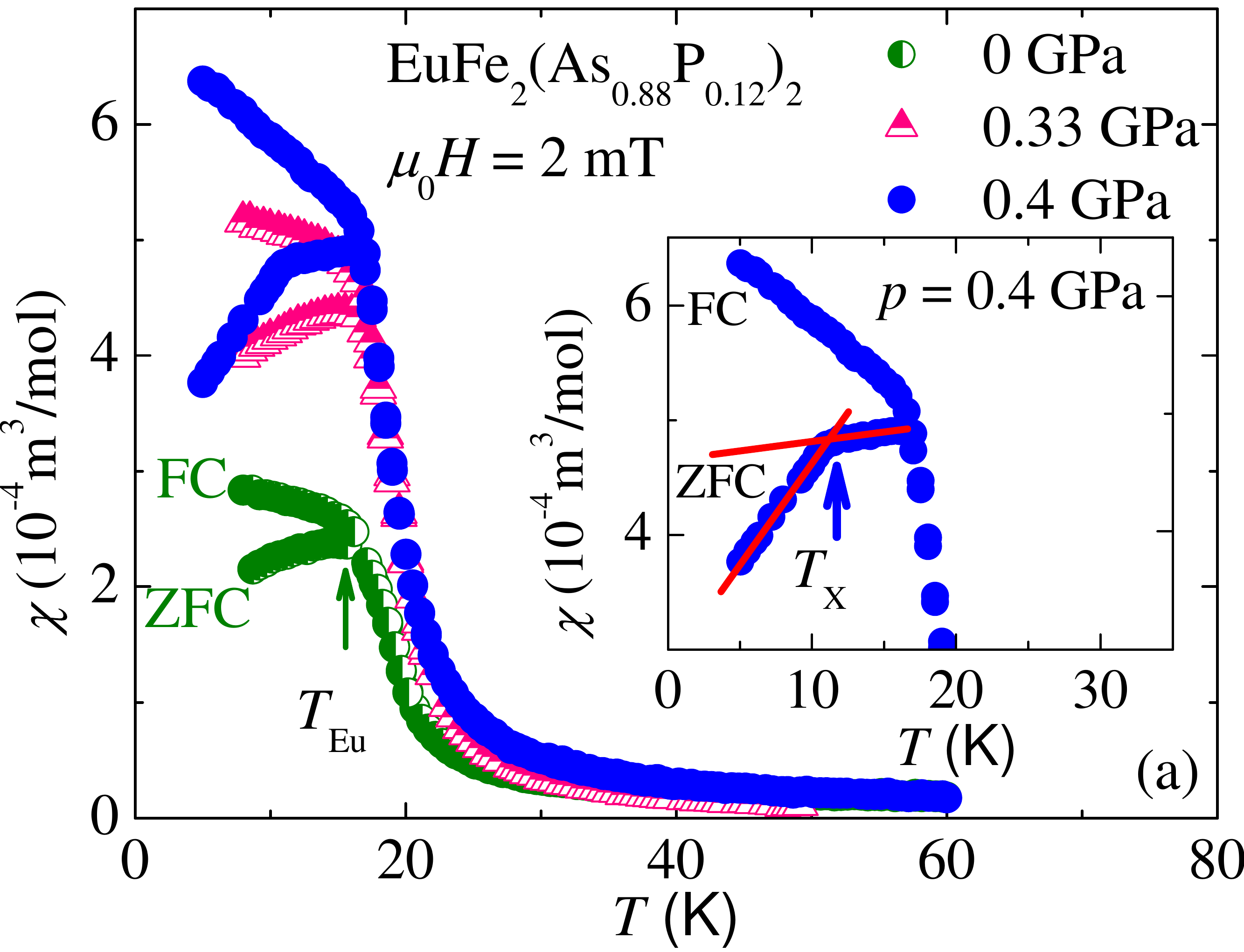}
\includegraphics[width=0.9\linewidth]{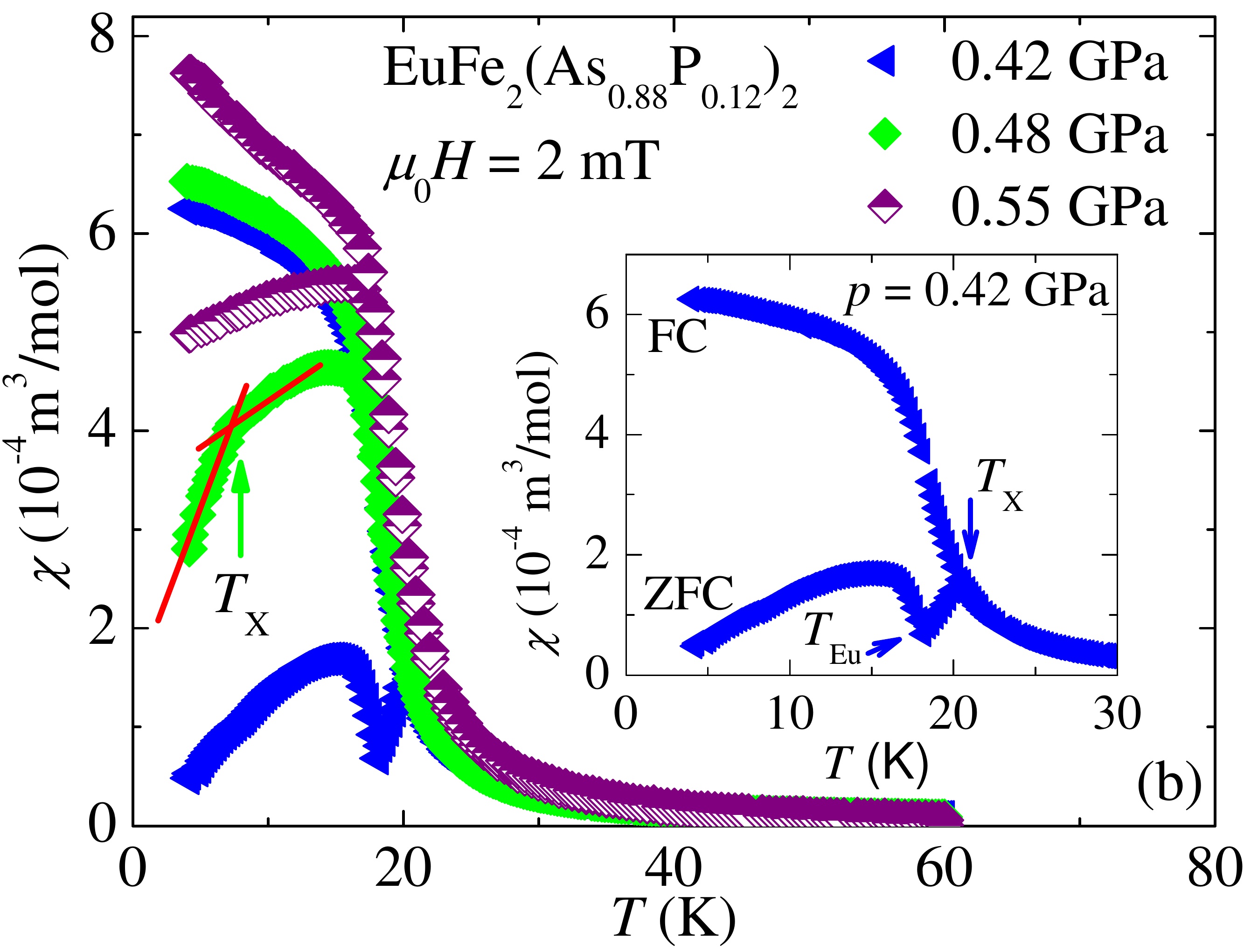}
\includegraphics[width=0.9\linewidth]{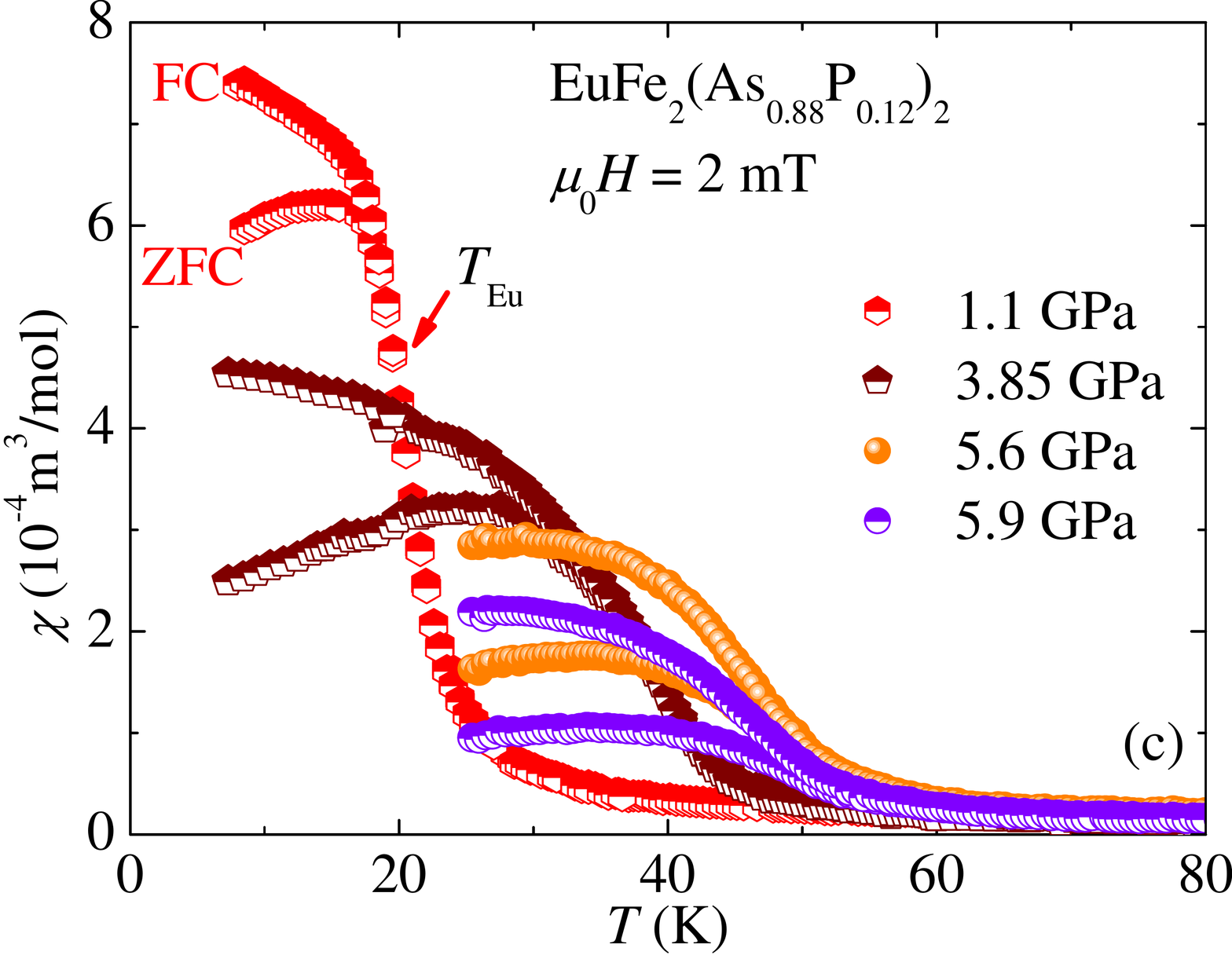}
\vspace{-0.3cm}
\caption{ (Color online) Temperature dependence of the ZFC and FC 
magnetic susceptibility of EuFe$_{2}$(As$_{0.88}$P$_{0.12}$)$_{2}$ in a magnetic field of ${\mu_{\rm 0}}$$H$ = 2~mT 
for $p$ ${\leq}$  0.4 GPa (a), for 0.42 GPa ${\leq}$ $p$ ${\leq}$ 0.55 GPa (b), and 
for 1.1 GPa ${\leq}$ $p$ ${\leq}$ 5.9 GPa (c).
The arrows mark the ordering temperature of Eu moments ($T_{\rm Eu}$) and the "X" transition temperatures ($T_{\rm X}$). 
The insets of panel (a) and (b) show the low temperature data for $p$ = 0.4 GPa and 0.42 GPa,
respectively, illustrating the transition
to the superconducting state marked by the arrows. The solid lines are guides to the eye.}
\label{fig1}
\end{figure}
 Magnetization measurements were carried out under hydrostatic pressures
up to $p$ = 5.9 GPa. The temperature dependence of the ZFC and FC  
magnetic susceptibilities ${\chi}$ for EuFe$_{2}$(As$_{0.88}$P$_{0.12}$)$_{2}$
recorded at ambient and selected applied 
pressures is shown in Fig.~6a ($p$ ${\leq}$  0.4 GPa), in Fig.~6b (0.42 GPa ${\leq}$ $p$ ${\leq}$ 0.55 GPa),
and in Fig.~6c (1.1 GPa ${\leq}$ $p$ ${\leq}$ 5.9 GPa). Note that Fig.~6 shows the data
after subtraction of the background signal 
from the empty pressure cell.     
The magnetic ordering temperature $T_{\rm Eu}$  
of the Eu$^{2+}$ moments was determined as described in Sec.~IIIA.2.
At ambient pressure a clear bifurcation between the ZFC and FC curves
appears below $T_{\rm Eu}$ ${\simeq}$ 16.5 K, which is consistent
with the susceptibility data obtained for the sample without pressure cell (see Fig.~2).
In addition, the magnitudes of the susceptibilities are also in fair agreement.
Upon increasing the pressure an anomaly in the ZFC susceptibility is observed at 
$p$ = 0.4 GPa, 0.42 GPa and 0.48 GPa as shown in Figs.~6a and 6b.
The low-temperature data for $p$ = 0.4 GPa are shown in the inset of Fig. 6a. In addition to the Eu order
observed at ${\simeq}$ 18 K, a strong decrease of the ZFC susceptibility is observed at 
${\simeq}$ 11 K, which is possibly due to the appearance of superconductivity.
The decrease of the susceptibility corresponds to nearly 100 ${\%}$ diamagnetic shielding.
In order to confirm superconductivity transport measurements under pressure are necessary. Only
magnetization data do not allow to conlcude that the observed decrease of ${\chi}_{\rm ZFC}$
is due to the appearance of  superconductivity. Hence we call this phase "X".  
For $p$ = 0.42 GPa the susceptibility also shows a pronounced decrease at $T_{\rm X}$ ${\simeq}$ 20 K (see Fig.~2b,
the low-temperature data are shown in the inset). 
Below ${\simeq}$ 18.2 K the susceptibility starts to increase again due to the C-AFM ordering of the Eu$^{2+}$ moments. 
Upon increasing the pressure to $p$ = 0.47 GPa the transition temperature $T_{\rm X}$ decreases to 8.8 K.  
Above $p$ = 0.55 GPa the "X" phase is no longer visible (see Fig.~6c). It is also absent for 
$p$ ${\textless}$ 0.35 GPa. Therefore, pressure-induced "X" phase in EuFe$_{2}$(As$_{0.88}$P$_{0.12}$)$_{2}$ 
is very likely present in a very narrow pressure range.
  We observed that $T_{\rm Eu}$ increases upon increasing
hydrostatic pressure, similar to chemical pressure. 
The maximum value of $T_{\rm Eu}$ = 56 K is reached at $p$ = 5.6 GPa, and
at higher pressures it tends to decrease (see Fig.~6c). 
For instance, $T_{\rm Eu}$ = 53 K at the maximum applied pressure $p$ = 5.9 GPa. 
A maximum of $T_{\rm Eu}$ was also observed for the parent compound 
EuFe$_{2}$As$_{2}$, but at higher pressure ($p$ = 8 GPa).
According to recent X-ray diffraction studies \cite{Uhoya} of 
EuFe$_{2}$As$_{2}$, a collapsed tetragonal
(cT) phase was found above 8 GPa. 
It is known that the pressure-induced structural transition toward the cT phase is
connected with a valence change of the Eu ions, as reported
for EuFe$_{2}$P$_{2}$ and EuCo$_{2}$P$_{2}$.\cite{Ni} Therefore, it is possible that 
the decrease of $T_{\rm Eu}$ above 5.6 GPa is connected with a pressure-induced valence change
from the magnetic Eu$^{2+}$ to the nonmagnetic Eu$^{3+}$
state. However, to gain further insight into this pressure region,
measurements at $p$ ${\textgreater}$ 5.9 GPa are necessary.\\  

\subsubsection{Zero-field ${\mu}$SR measurements under pressure}
\begin{figure*}[ht!]
\centering
\includegraphics[width=0.8\linewidth]{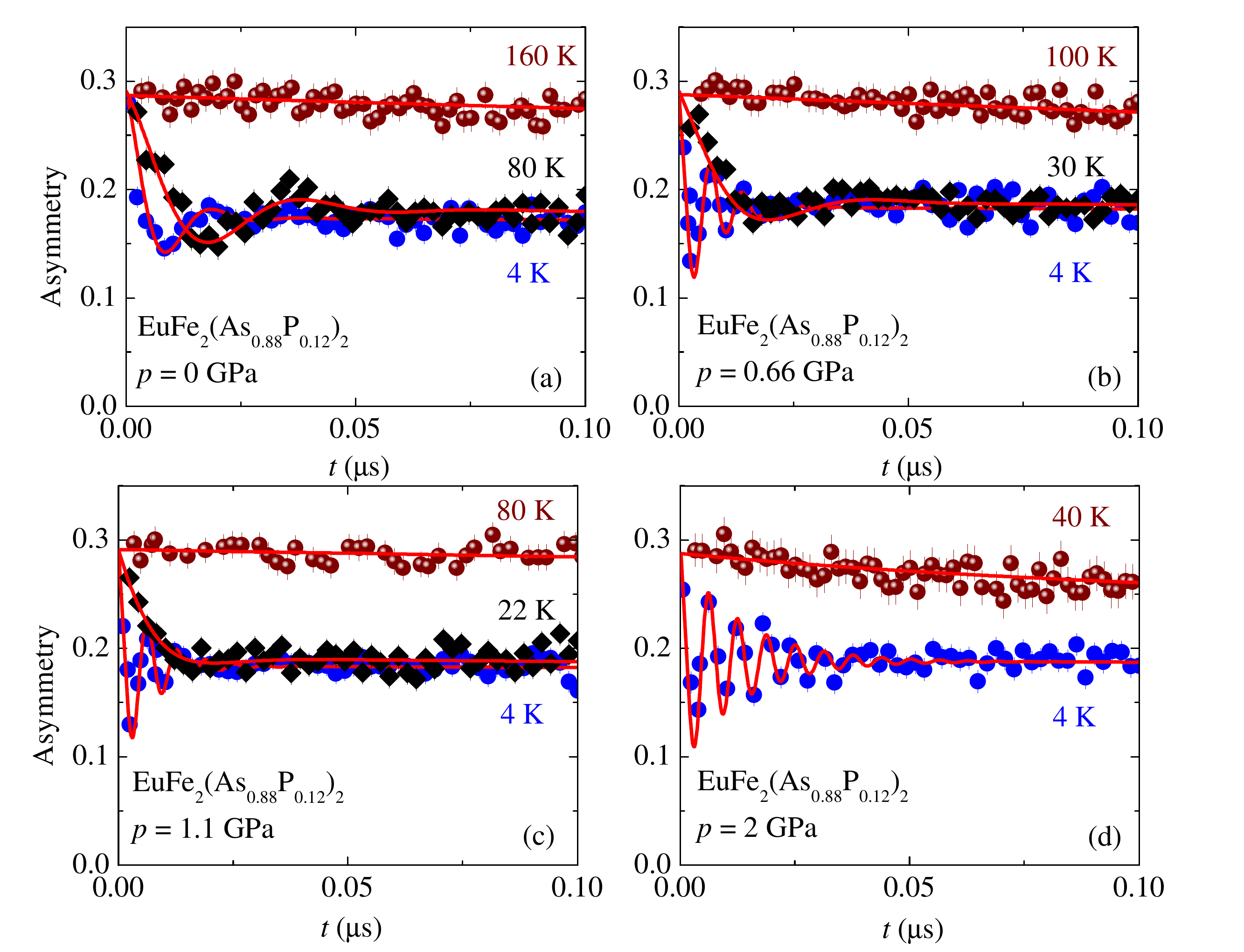}
\vspace{0cm}
\caption{ (Color online) ZF ${\mu}$SR spectra of EuFe$_{2}$(As$_{0.88}$P$_{0.12}$)$_{2}$ 
measured at $p$ = 0.0, 0.66, 1.1 and 2 GPa, recorded for three different temperatures: 
$T$ ${\textless}$ $T_{\rm Eu}$ (circles), $T_{\rm Eu}$ ${\textless}$ $T$ ${\textless}$ $T_{\rm SDW}$ (diamonds), 
and $T$ ${\textgreater}$ $T_{\rm SDW}$ (spheres). 
The solid lines represent fits to the data by means of Eq.~(4).}
\label{fig3} 
\end{figure*}
 Hydrostatic pressure effects on the magnetic properties of  
EuFe$_{2}$(As$_{0.88}$P$_{0.12}$)$_{2}$ were studied microscopically
by means of ZF ${\mu}$SR. Some representative ${\mu}$SR time spectra 
at different applied pressures are shown in Fig.~7. 
A substantial fraction of the ${\mu}$SR asymmetry signal originates 
from muons stopping in the MP35N pressure cell \cite{Andreica} surrounding the sample. 
Therefore, the total ${\mu}$SR asymmetry is a sum of two components: 
\begin{equation}
A^{ZF}(t)=A_S^{ZF}(t)+A_{PC}^{ZF}(t),
\end{equation}
$A_{S}^{ZF}$(t) is the contribution of the sample, and $A_{PC}^{ZF}$(t) is 
the contribution of the pressure cell. $A_{S}^{ZF}$(t) is well described by 
Eq.~(2) with ${\alpha}_{\rm 1}$ = 2/3 and ${\beta}_{\rm 1}$ = 1/3 
(since for $x$ = 0.12 the ${\mu}$SR spectra contain only one frequency, ${\alpha}_{\rm 2}$ = 0 and ${\beta}_{\rm 2}$ = 0). 
The signal of the pressure cell was analyzed by a damped Kubo-Toyabe (KT) function:\cite{Andreica}
\begin{equation}
A_{PC}^{ZF}(t)=A_{PC}^{ZF}(0)\Bigg[\frac{1}{3}+\frac{2}{3}(1-{\sigma}t)e^{{-\sigma^2}{t^2}/{2}}\Bigg]e^{-{\lambda}t}.
\end{equation}
Here $A_{PC}^{ZF}$(0) is the amplitude of $A_{PC}^{ZF}$(t) at t = 0.
The width of the static Gaussian field distribution ${\sigma}$ = 0.338 ${\mu}$$s^{-1}$
and the damping rate ${\lambda}$ = 0.04 ${\mu}$$s^{-1}$ were obtained from a measurement
of the empty pressure cell.  
The total initial asymmetry is $A_{S}^{ZF}$(0)+$A_{PC}^{ZF}$(0) = 0.29. 
The ratio $A_{S}^{ZF}$(0)/[$A_{S}^{ZF}$(0)+$A_{PC}^{ZF}$(0)] ${\simeq}$ 40${\%}$
implies that approximately 40${\%}$ of the muons are stopping in the sample.
\begin{figure}[b!]
\includegraphics[width=0.98\linewidth]{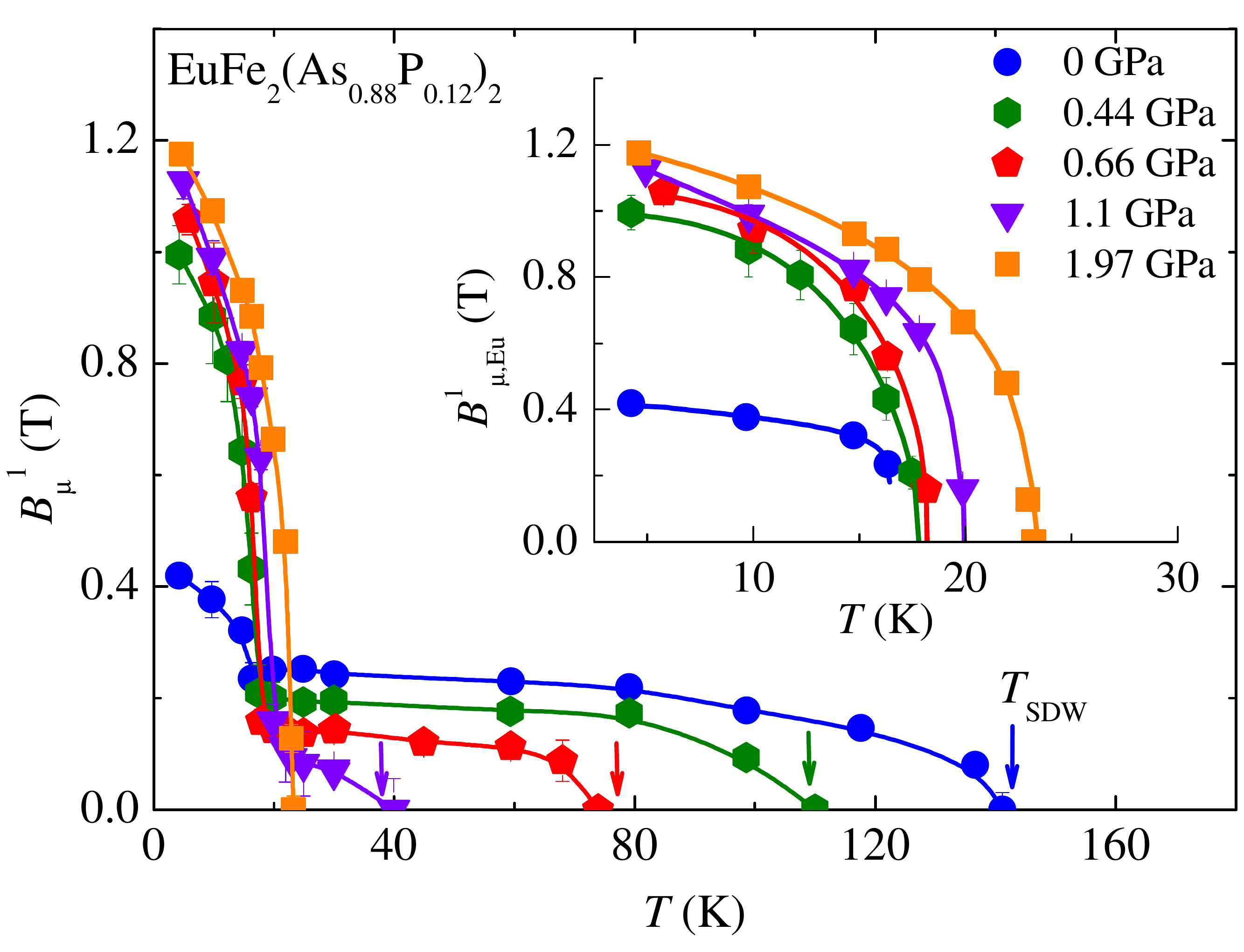}
\vspace{-0.3cm}
\caption{ (Color online) Temperature dependence of the internal field $B^1_{\rm \mu}$ 
at the muon site for the sample EuFe$_{2}$(As$_{0.88}$P$_{0.12}$)$_{2}$ 
recorded at various applied pressures. 
The solid lines represent fits to the data by means of Eq.~(3). The arrows mark the ordering  
temperature $T_{\rm SDW}$. The inset shows the low temperature data, illustrating the
transition at $T_{\rm Eu}$ to the magnetically ordered state of the Eu moments.}
\label{fig1}
\end{figure}
\begin{table*}[t!]
\caption{Summary of the parameters obtained for the polycrystalline sample of EuFe$_{2}$(As$_{0.88}$P$_{0.12}$)$_{2}$ 
at different hydrostatic pressures by means of magnetization and ${\mu}$SR experiments. The meaning of the
symbols is given in the text.}
\vspace{0.3cm}
\begin{tabular}{lccccccc}
\hline
\hline
    & $p$        &$T_{\rm Eu}^{\chi}$ & $T_{\rm Eu}^{{\mu}SR}$ & $T_{\rm c}$ & $T_{\rm SDW}^{{\mu}SR}$ &  $B^1_{\rm \mu,Eu}$ (0) &  $B^1_{\rm \mu,SDW}$ (0) \\
    & (GPa)          & (K)   & (K)   & (K)   & (K)   & (T)   & (T) \\ \hline
    & 0                   & 16.5(5) & 16.4(3) & - & 141.2(1) & 0.44(1) & 0.25(1) \\ 
     & 0.4                 &  17.5(4)  &    -  & 11.2(3)   & -      & -       &    -       \\
     & 0.42                &  17.6(5)  &    -  & 20.3(3)   & -      & -        &   -        \\
    & 0.44                &    -       & 17.7(3) & - & 110(1) & 0.99(2) & 0.19(2)\\
     & 0.48               &   17.9(4)  & -      & 7.3(3)   & - & - & -            \\
     & 0.66                & 18.9(5) & 18.4(6) & 0 & 75(2) & 1.07(5) & 0.17(4)\\ 
     & 1.1                 & 20.5(5) & 19.9(7) & 0 & 40(3) & 1.27(3) & 0.12(2)\\   
     & 1.73                & 24.6(3) & - & 0 & - & - & - \\               
     & 1.97                & 24.4(5) & 23.6(7) & 0  & 0 & 1.23(2) & 0\\
     & 2.5                 & 27.2(3) & - & 0  & - & - & - \\                   
     & 3.85                & 35.8(5) & - & 0  & - & - & - \\ 
     & 4.54                & 42.5(4) & - & 0 & - & - & - \\  
     & 5.1                 & 49.5(3) & - & 0 & - & - & - \\                
     & 5.6                 & 57(4) & -  & 0 & - & - & - \\                  
     & 5.9                 & 53(5) & - & 0 & - & - & - \\                   \hline
\hline         
\end{tabular}
\label{table1}
\end{table*}
 Up to $p$ = 1.1~GPa the spontaneous muon-spin precession 
in the Eu ordered and in the SDW state is clearly observed in the ZF ${\mu}$SR time spectra (see Fig.~7), 
indicating long range magnetic order in the Eu and the Fe sublattice.
Above $p$ = 1.1 GPa the SDW state is suppressed and only the magnetic order of the Eu moments remains.
The temperature dependence of the internal field $B^1_{\rm \mu}$ 
for various hydrostatic pressures is shown in Fig.~8. 
The inset shows $B^1_{\rm \mu,Eu}$ at low temperatures where the magnetic ordering of 
the Eu$^{2+}$ moments is evident. The data were analyzed by Eq.~(3).
The SDW ordering temperature $T_{\rm SDW}$ of the Fe moments ($T_{\rm SDW}$ = 140 K at 
ambient pressure) as well as $B^1_{\rm \mu,SDW}$ created by the Fe sublattice decrease with increasing pressure.
Above $p$ = 1.1 GPa the SDW order is completely suppressed.
On the contrary, $T_{\rm Eu}$ increases with pressure, in agreement with the susceptibility measurements.
In addition, $B^1_{\rm \mu,Eu}$ related to the Eu ordered state also 
increases with pressure. Note the sharp increase of $B^1_{\rm \mu, Eu}$ below $T_{\rm Eu}$ with
increasing the pressure from $p$ = 0 GPa to $p$ = 0.44 GPa. For $p$ ${\textgreater}$ 0.44 GPa a more 
smooth increase of $B^1_{\rm \mu, Eu}$ is observed. 
Relevant parameters of EuFe$_{2}$(As$_{0.88}$P$_{0.12}$)$_{2}$
extracted from the high pressure magnetization and ${\mu}$SR experiments 
are listed in Table~II.

 As shown above the magnetization measurements indicate the presence 
of a possible SC phase in the pressure range 0.36 GPa ${\leq}$ $p$ ${\leq}$ 0.5 GPa.
An attempt to detect it in the sample EuFe$_{2}$(As$_{0.88}$P$_{0.12}$)$_{2}$
with ZF and TF ${\mu}$SR, failed because of the strong intrinsic magnetism present in the sample. 

\section{Phase diagram}
 
 Fig.~9a shows the ($x$-$T$) phase diagram for the system EuFe$_{2}$(As$_{1-x}$P$_{x}$)$_{2}$.
The ($p$-$T$) phase diagram of EuFe$_{2}$(As$_{0.88}$P$_{0.12}$)$_{2}$ is plotted in
Fig.~9b. The data for $T_{\rm Eu}$ represented by the triangles
in Fig.~9a are taken from Ref.~25. 
In the ($x$-$T$) phase diagram 
three different phases were identified: 
a paramagnetic phase (PM), spin-density wave order
of the Fe moments (SDW), and magnetic order of Eu$^{2+}$ moments (MO).  
Moreover, in the ($p$-$T$) phase diagram, pressure-induced "X" phase was found (see inset of Fig.~9b). 
In Fig.~10 the internal magnetic fields $B^1_{\rm \mu,Eu}$ and $B^1_{\rm \mu,SDW}$ 
probed by the muons in the Eu ordered and in the SDW state and the low temperature value of the 
magnetic susceptibility ${\chi}_{\rm ZFC}$(7 K) are plotted as a function of P content $x$ and applied pressure $p$.
\begin{figure}[t!]
\includegraphics[width=1.0\linewidth]{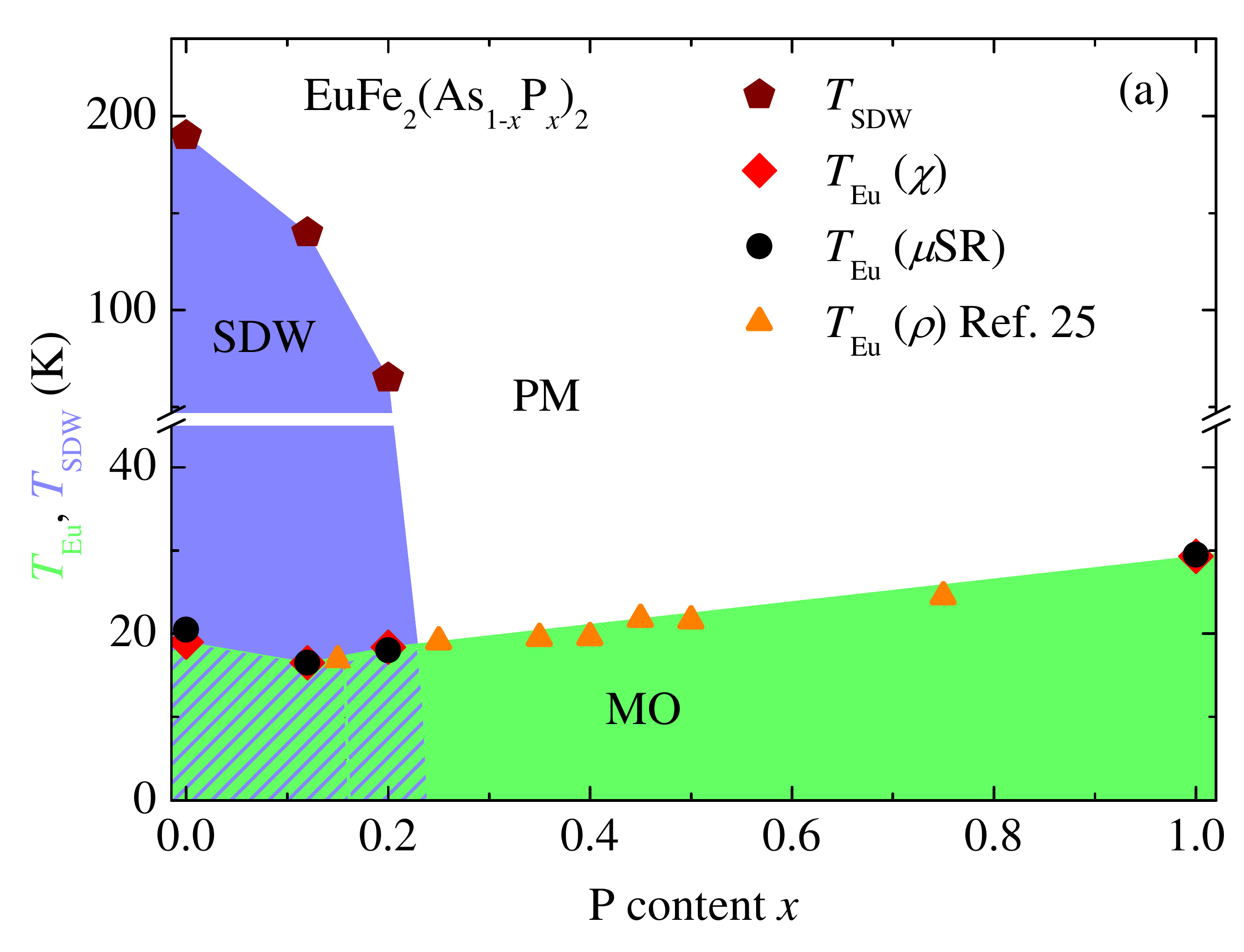}
\includegraphics[width=1.0\linewidth]{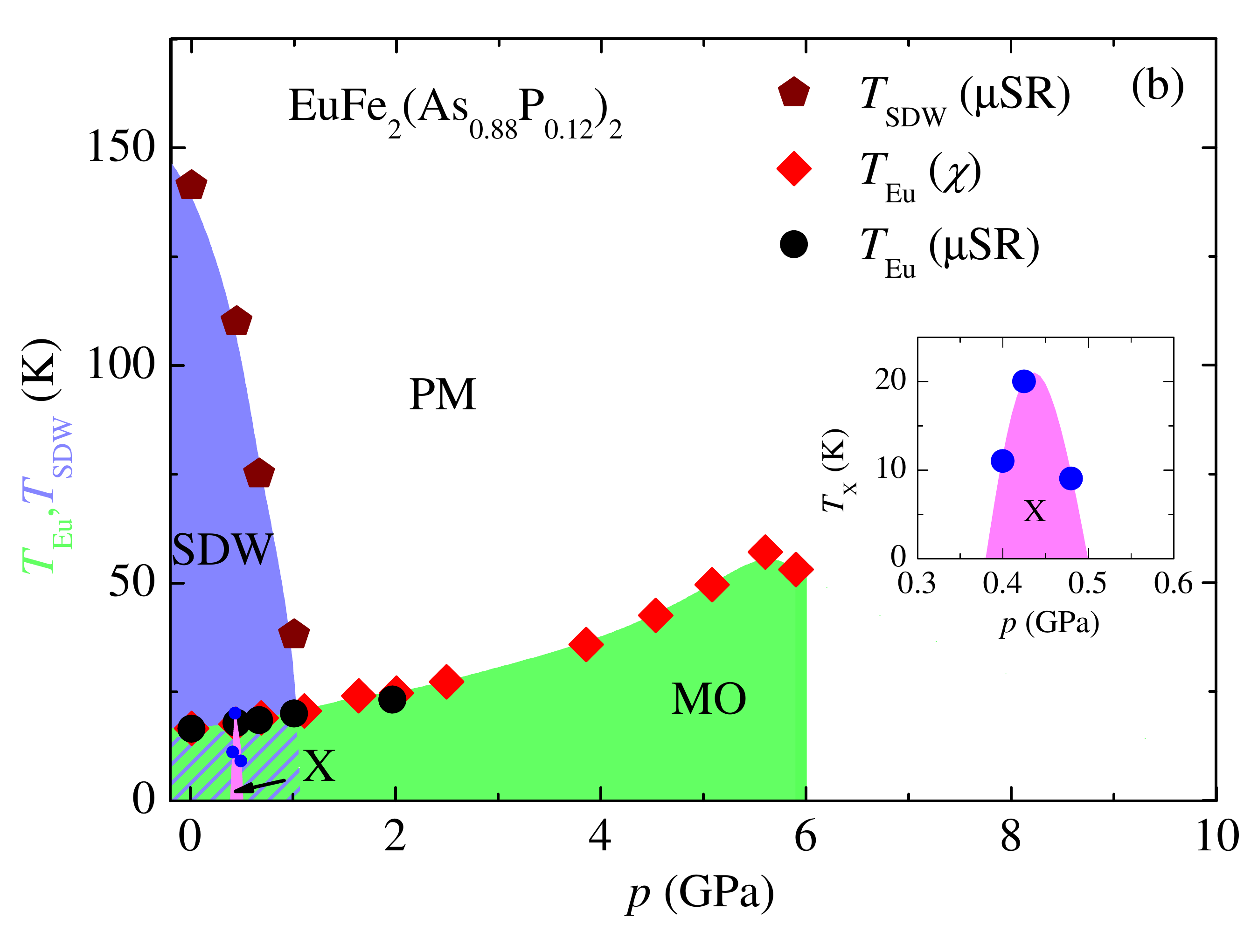}
\vspace{-0.8cm}
\caption{ (Color online) (a) ($x$-$T$) phase diagram of EuFe$_{2}$(As$_{1-x}$P$_{x}$)$_{2}$.
The data points represented by the triangles are taken from Ref.~25.
(b) ($p$-$T$) phase diagram of EuFe$_{2}$(As$_{0.88}$P$_{0.12}$)$_{2}$.  
The various phases in the phase diagrams  and the corresponding transition temperatures are denoted as follows: 
paramagnetic (PM), spin-density wave (SDW) and $T_{\rm SDW}$,
magnetic ordering of Eu (MO) and $T_{\rm Eu}$, "X" phase (the meaning of this phase is given in the text) and $T_{\rm X}$.
For clarity the inset in (b) shows the "X" phase present in a very narrow pressure range.} 
\label{fig1}
\end{figure}
\begin{figure}[t!]
\includegraphics[width=1.7\linewidth]{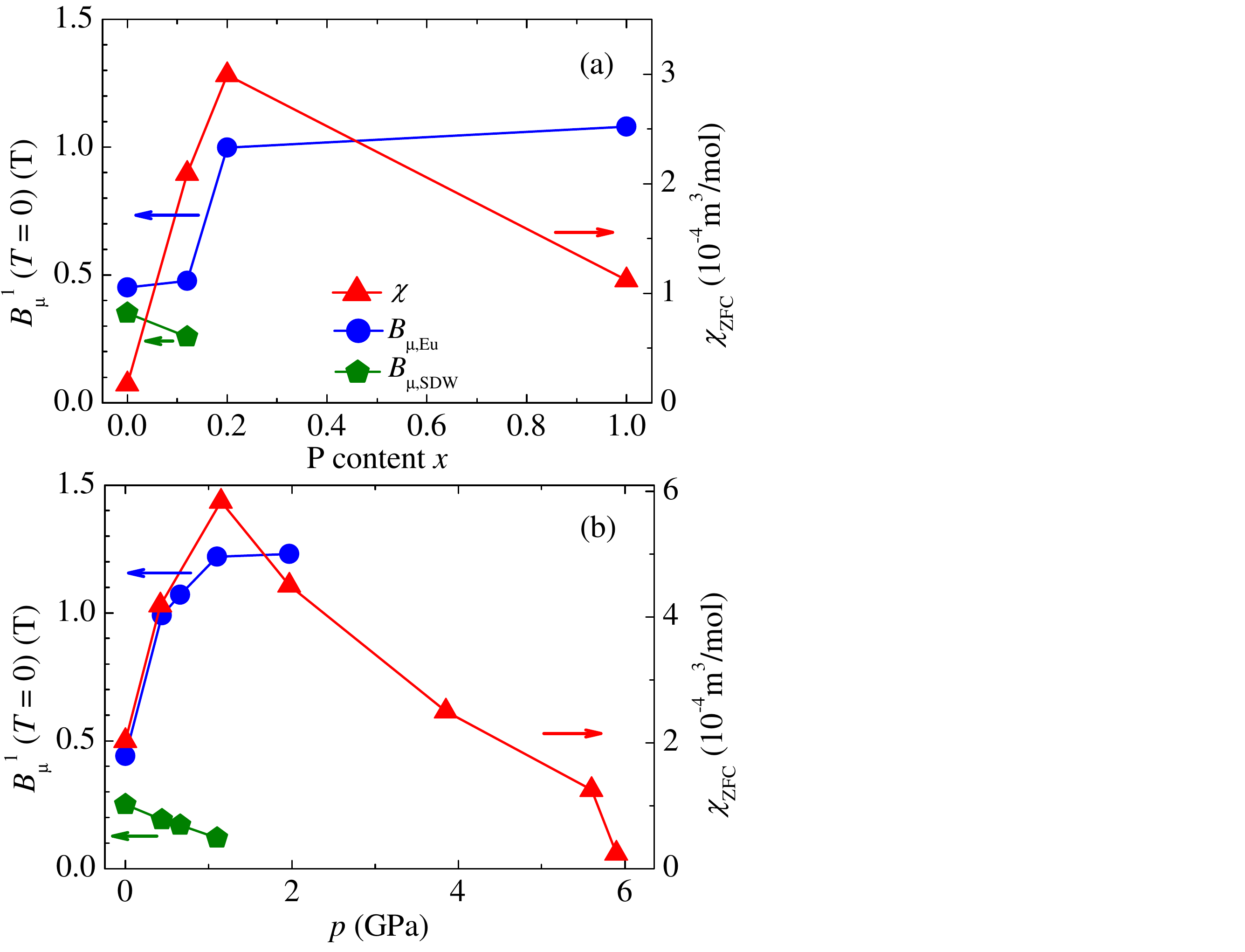}
\vspace{-0.5cm}
\caption{ (Color online) Zero-temperature values of the internal magnetic fields
$B^1_{\rm \mu,Eu}$ and $B^1_{\rm \mu,SDW}$ 
and the low-temperature value of the magnetic susceptibility ${\chi}_{\rm ZFC}$ as a function of 
the P content $x$ (a) and applied pressure (b).}
\label{fig1}
\end{figure}

  By combining the above phase diagrams one obtains
a coherent physical picture on the system 
EuFe$_{2}$As$_{2}$ upon P substitution and on EuFe$_{2}$(As$_{0.88}$P$_{0.12}$)$_{2}$ under hydrostatic pressure. 
An important finding is the observation of pressure-induced superconductivity in 
EuFe$_{2}$(As$_{0.88}$P$_{0.12}$)$_{2}$, 
coexisting with magnetic order of the Eu and Fe moments. 
Superconductivity appears in the narrow pressure region of 0.36-0.5 GPa.  
The presented phase diagrams in combination
with the results obtained for the parent compound under pressure \cite{Terashima,Matsubayashi} allow us to
draw the following conclusion on the relation between chemical and hydrostatic pressure
in EuFe$_{2}$As$_{2}$: 

1) Both chemical and hydrostatic pressure suppress $T_{\rm SDW}$ and $B^1_{\rm \mu,SDW}$(0). 
However, the SDW ground state is differently affected by $x$ and $p$.
At all applied pressures below $p$ = 1.1~GPa 
long-range SDW order was observed, while in the case of chemical pressure for $x$ = 0.2 
a disordered SDW phase exist. This may be related to the fact that by chemical pressure (P substitution) 
considerably more disorder is introduced.

2) Fig.~9 shows that in the case of P substitution
$T_{\rm Eu}$ first decreases as a function of $x$, reaches a minimum at $x$ = 0.12,
and then increases. For a fixed P content of $x$ = 0.12 the ordering temperature $T_{\rm Eu}$
increases with pressures up to $p$ = 5.6 GPa. Above $p$ = 5.6 GPa, however,
$T_{\rm Eu}$($p$) decreases, acompanied by a possible valence change of the Eu moments. 
In the parent compound EuFe$_{2}$As$_{2}$ a valence change was found at a higher pressure $p$ = 8 GPa.

3) The internal magnetic field $B^1_{\rm \mu,Eu}$(0) in the Eu ordered state 
increases with increasing $x$ as well as by applying hydrostatic pressure (see Fig.~10a and b).

4) The low temperature value of the magnetic susceptibility ${\chi}_{\rm ZFC}$ (7 K) first increases
with increasing $x$ and $p$ and above some critical values 
($x$ = 0.2 and $p$ = 1.1 GPa) it decreases (see Figs.~10a and b).


 By considering the findings listed above the qualitative statement can be made
that the properties of EuFe$_{2}$(As$_{1-x}$P$_{x}$)$_{2}$
are similarly tuned by chemical and hydrostatic pressure.

\section{CONCLUSIONS}
In summary, the magnetic and superconducting properties of the
system EuFe$_{2}$(As$_{1-x}$P$_{x}$)$_{2}$ ($x$ = 0, 0.12, 0.2, 1) 
were studied by magnetization and ${\mu}$SR experiments.
In addition, the sample with $x$ = 0.12 was also investigated
by applying hydostatic pressure up to $p$ ${\simeq}$ 5.9 GPa. The
($x$-$T$) phase diagram of EuFe$_{2}$(As$_{1-x}$P$_{x}$)$_{2}$ and
the ($p$-$T$) phase diagram of EuFe$_{2}$(As$_{0.88}$P$_{0.12}$)$_{2}$
were determined and discussed as well as compared to the ($p$-$T$) 
phase diagram recently obtained for EuFe$_{2}$As$_{2}$.\cite{Terashima,Matsubayashi}  
The present investigations reveal that the magnetic coupling between the Eu and 
the Fe sublattices strongly depends on chemical and hydrostatic pressure and determines the
($x$-$T$) and ($p$-$T$) phase diagrams as presented in this work.
According to the above discussed phase diagrams chemical and hydrostatic pressures
have qualitatively a similar effect on the Fe and Eu magnetic order.

 There are still some open questions related to superconductivity and its interplay
with the magnetic ground state of the system EuFe$_{2}$As$_{2}$. One of the most interesting aspects of 
this particular member of Fe-based superconductors
is the possibility to observe coexistence or competition between superconductivity and
rare-earth Eu magnetic order. In the present work the so called "X" phase induced by pressure was 
observed in  EuFe$_{2}$(As$_{0.88}$P$_{0.12}$)$_{2}$ in addition to the magnetic phases of the Eu and Fe sublattices. 
It exists in a narrow pressure range 0.36-0.5 GPa. This phase is possibly superconducting.
However, transport measurements as a function 
of pressure are required in order to clarify this point.

\section{Acknowledgments}~
 ~This work was supported by the Swiss National Science Foundation, the
SCOPES grant No. IZ73Z0${\_}$128242, and the NCCR Project MaNEP.
The ${\mu}$SR experiments were performed at the Swiss Muon Source
of the Paul Scherrer Institute (PSI), Villigen, Switzerland.

\end{document}